\documentclass[11pt]{article}
\usepackage{amsfonts,amsmath,amssymb,amsthm,textcomp}
\usepackage{epic}
\usepackage{graphicx}
\usepackage{epsfig}
\usepackage{latexsym}

\allowdisplaybreaks
\numberwithin{equation}{section}

\newtheorem{thm}{Theorem}[section]
\newtheorem{lem}[thm]{Lemma}
\newtheorem{cor}[thm]{Corollary}
\newtheorem{prop}[thm]{Proposition}
\newtheorem{deff}[thm]{Definition}
\newtheorem{conj}[thm]{Conjecture}
\newtheorem{condition}[thm]{Condition}

\theoremstyle{definition}
\newtheorem{example}[thm]{Example}
\newtheorem{rem}[thm]{Remark}

\newcommand{\bconj}{\begin{conj}}
\newcommand{\econj}{\end{conj}}
\newcommand{\bth}{\begin{thm}}
\newcommand{\ethGL}{\end{thm}}
\newcommand{\bl}{\begin{lem}}
\newcommand{\el}{\end{lem}}
\newcommand{\bdf}{\begin{deff}}
\newcommand{\edf}{\end{deff}}
\newcommand{\bcor}{\begin{cor}}
\newcommand{\ecor}{\end{cor}}
\newcommand{\bprop}{\begin{prop}}
\newcommand{\eprop}{\end{prop}}
\newcommand{\brem}{\begin{rem}}
\newcommand{\erem}{\end{rem}}
\newcommand{\beq}{\begin{equation}}
\newcommand{\eeq}{\end{equation}}
\newcommand{\beqn}{\begin{eqnarray}}
\newcommand{\eeqn}{\end{eqnarray}}
\newcommand{\beqns}{\begin{eqnarray*}}
\newcommand{\eeqns}{\end{eqnarray*}}

\newenvironment{romenumerate}{\begin{enumerate}
 }{\end{enumerate}}

\newcommand{\bpr}{\begin{proof}}
\newcommand{\epr}{\end{proof}}

\newcommand{\BO}{\mathcal{O}}
\newcommand{\bo}{o}

\newcommand{\ba}{\begin{array}}
\newcommand{\ea}{\end{array}}
\newcommand{\bit}{\begin{itemize}}
\newcommand{\eit}{\end{itemize}}
\newcommand{\ben}{\begin{enumerate}}
\newcommand{\een}{\end{enumerate}}
\newcommand{\bal}{\begin{align}}
\newcommand{\bals}{\begin{align*}}
\newcommand{\bs}{\begin{skip}}
\newcommand{\eal}{\end{align}}
\newcommand{\eals}{\end{align*}}
\newcommand{\es}{\end{skip}}
\newcommand{\babs}{\begin{abstract}}
\newcommand{\eabs}{\end{abstract}}

\newcommand{\lp}{\left (}
\newcommand{\rp}{\right )}
\newcommand{\kl}{[\hspace{-0.5 mm}[}
\newcommand{\kr}{]\hspace{-0.5 mm}]}

\def\ii{\mathbf{i}}

\newcommand{\tX}{\widetilde{X}}

\newcommand{\de}{\delta}

\newcommand{\La}{\Lambda}
\newcommand{\al}{\alpha}
\newcommand{\be}{\beta}
\newcommand{\ra}{\rightarrow}

\newcommand{\eps}{\varepsilon}
\newcommand{\II}{\infty}

\newcommand{\gam}{\gamma}

\def\a{\alpha}

\renewcommand{\Pr}{\mathbb{P}}

\newcommand{\E}{\mbox{$\mathbb E$}}
\newcommand{\M}{\mbox{$\mathbb M$}}
\newcommand{\V}{\mbox{$\mathbb V$}}
\def\P{{\mathbb {P}}}

\def\a{\alpha}

\newcommand{\refT}[1]{Theorem~\ref{#1}}

\newcommand{\refR}[1]{Remark~\ref{#1}}
\newcommand{\refS}[1]{Section~\ref{#1}}

\newcommand{\refE}[1]{Example~\ref{#1}}
\newcommand{\refCC}{Condition~\ref{C1}}
\newcommand{\refand}[2]{\ref{#1} and~\ref{#2}}

\newcommand\set[1]{\ensuremath{\{#1\}}}

\newcommand\bigpar[1]{\bigl(#1\bigr)}
\newcommand\Bigpar[1]{\Bigl(#1\Bigr)}

\newcommand\bigsqpar[1]{\bigl[#1\bigr]}

\newcommand\dto{\overset{\mathrm{d}}{\to}}

\newcommand\eqd{\overset{\mathrm{d}}{=}}
\renewcommand\={:=}

\newcommand\bbR{\mathbb R}

\newcommand\bbZ{\mathbb Z}

\newcounter{CC}
\newcommand{\CC}{\stepcounter{CC}\CCx} 
\newcommand{\CCx}{C_{\arabic{CC}}}     
\newcounter{cc}

\newcommand\ie{i.e.\spacefactor=1000}
\newcommand\eg{e.g.\spacefactor=1000}
\newcommand{\as}{a.s.\spacefactor=1000}

\newcommand{\cL}{\mathcal{L}}
\newcommand{\cP}{\mathcal{P}}

\newcommand\Bi{\operatorname{Bi}}
\newcommand\Be{\operatorname{Be}}
\newcommand\Ge{\operatorname{Ge}}
\newcommand\gb{\beta}
\newcommand\gd{\delta}
\newcommand\loga{\log_{1/\alpha}}
\newcommand\nn{{n\ge2}}
\newcommand\ett{\mathbf 1}
\newcommand\qw{^{-1}}
\newcommand\floor[1]{\lfloor#1\rfloor}
\newcommand\ceil[1]{\lceil#1\rceil}
\newcommand\dw{d_\mathrm{W}}
\newcommand\dtv{d_\mathrm{TV}}
\newcommand\ntoo{\ensuremath{{n\to\infty}}}
\newcommand\xxl{\mathrm{l}}
\newcommand\xxc{\mathrm{c}}
\newcommand\xl{_{\xxl}}
\newcommand\xc{_{\xxc}}
\newcommand\DF{\Delta F}
\newcommand\tpsi{\tilde\psi}

\title{\bf Convergence of some leader election algorithms}
\author {
Svante Janson\thanks{Uppsala University, Department of Mathematics, PO
 Box 480, SE-751 06 Uppsala,  Sweden. 
\texttt{svante.janson@math.uu.se}} 
\and 
Christian Lavault\thanks{ LIPN (UMR CNRS 7030),\ Universit\'e Paris 13,\
99, av. J.-B. Cl\'ement 93430 Villetaneuse,\ France.
{\tt lavault@lipn.univ-paris13.fr }} 
\and
Guy Louchard\thanks{Universit\'e Libre de Bruxelles,
D\'epartement d'Informatique, CP 212, Boulevard du Triomphe, B-1050
Bruxelles, Belgium. {\tt louchard@ulb.ac.be}}
}

\date{February 8, 2008} 
\begin{document}

\maketitle

\babs 
We start with a set of $n$ players. With some probability $P(n,k)$, we
kill $n-k$ players; the other ones stay alive, and we repeat with them.
What is the distribution of the number $X_n$ of \emph{phases} (or
rounds) before getting only one player?  We present a  probabilistic
analysis of this algorithm under some conditions on the probability
distributions $P(n,k)$, 
including stochastic monotonicity and
the assumption that roughly a fixed proportion $\al$ of the
players survive in each round.

We prove a kind of convergence in distribution for $X_n-\loga n$; 
as in many other similar problems
there are oscillations and no true limit distribution, but suitable
subsequences converge, and there is an absolutely continuous random variable
$Z$ such that 
$d(X_n, \lceil Z+\loga n\rceil)\ra 0$, where $d$ is either the total
variation distance or the Wasserstein distance.

Applications of the general result include the leader election
algorithm where players are eliminated by independent coin tosses
and a variation of the
leader election algorithm proposed by W.R. Franklin \cite{FR82}.
We study the latter algorithm further, including numerical results.
\eabs

%

%
%
\section{A general convergence theorem}\label{SJ0}

We consider a general leader election algorithm of the following type:
We are given some random procedure that,  given any set of $n\ge2$
individuals, eliminates some (but not all) individuals. If there is
more that one survivor, we repeat the procedure with the set of
survivors until only one (the winner) remains. We are interested in
the (random) number $X_n$ of rounds required if we start with $n$
individuals.
(We set $X_1=0$, and have $X_n\ge1$ for $n\ge2$.)
We let $N_k$ be the number of individuals remaining after round $k$;
thus $X_n:=\min\set{k:N_k=1}$, where we start with $N_0=n$.
For convenience we may suppose that we continue with infinitely many
rounds where nothing happens; thus $N_k$ is defined for all $k\ge0$ and
$N_k=1$ for all $k\ge X_n$. 

We assume that the number $Y_n$ of survivors of a set of $n$
individuals has a distribution depending only on $n$. 
We have $1\le Y_n\le n$;
we allow the possibility that $Y_n=n$, but we assume $\Pr(Y_n=n)<1$
for every $n\ge2$, so that we will 
not get stuck before selecting a winner. 
We further
assume that, given the number of remaining individuals at the start of
a new round, the number of survivors is independent of the previous
history.
In other words, the sequence $(N_k)_0^\infty$ is a Markov chain
on $\{1,2,\ldots\}$, and
$X_n$ is the number of steps to absorption in $1$.
The transition probabilities of this Markov chain are,
with $Y_1=1$,
\begin{equation}
  P(i,j)\=\Pr(Y_i=j)=\Pr(\text{$j$ survives of a set of $i$}).
\end{equation}
Note that  $P(i,j)=0$ if $j>i$
and $P(i,i)<1$, for $i>1$. 
Conversely, any Markov chain on $\{1,2,\ldots\}$ with such $P(i,j)$
can be regarded as a leader election algorithm in the generality just
described.

We will in this paper treat leader election algorithms where,
asymptotically, a fixed proportion is eliminated in each round. (Thus,
we expect $X_n$ to be of the order $\log n$.)
More precisely, we assume the following for $Y_n$,
where we also repeat the key assumptions above.
(Here and below, $\log n$ should be interpreted as some fixed positive
number when $n=1$.)
\begin{condition}\label{C1}
For every $n\ge1$, $Y_n$ is a random variable such that
$1\le Y_n\le n$, and $\P(Y_n=n)<1$ for $n\ge2$. Further:
\begin{romenumerate}
\item\label{C1a}
$Y_n$ is stochastically increasing in $n$, \ie, 
$\Pr(Y_n\leq k)\geq \Pr(Y_{n+1}\leq k)$ for all $n\geq 1$ and $k\geq 1$. 
Equivalently, we may couple $Y_n$ and $Y_{n+1}$ such that $Y_n\leq Y_{n+1}$.

\item\label{C1b}
For some constants $\al\in(0,1)$ and $\eps>0$ and a sequence 
$\de_n=O\bigpar{(\log n)^{-1-\eps}}$,
\begin{equation}\label{c1b}
 \E Y_{n+1}-\E Y_n=\al+\BO(\de_n). 
\end{equation}

\item\label{C1c}
For some $\eps$ and $\gd_n$ as in \ref{C1b},
\begin{equation}\label{c1c}
\Pr(|Y_n-\al n|>\de_n n)=\BO(n^{-2-\eps}).
\end{equation}
\end{romenumerate}
\end{condition}

Note that 
\begin{equation}\label{ymom}
  \E|Y_n-\al n|^p=\BO(n^{p/2}) 
\end{equation}
for some $p>4$ suffices for \ref{C1c},
for a suitable choice of $\eps>0$ and $\gd_n$ (\eg, $\gd_n=n^{-\eta}$,
$\eta>0$ and $\eps$ small).

\begin{rem}\label{Rdelta}
If \eqref{c1b} or \eqref{c1c} holds for some sequence $(\gd_n)$, it holds
for every larger sequence $(\gd_n)$ too; similarly, if 
$\gd_n=O\bigpar{(\log n)^{-1-\eps}}$ or \eqref{c1c} 
holds for some $\eps$, it holds for every smaller $\eps$ too. Hence we
may assume that \ref{C1b} and \ref{C1c} hold with the same $\eps>0$
and the same $\gd_n$, and we may assume $\gd_n\ge(\log n)^{-1-\eps}$.
In particular, this implies that $\gd_k=\BO(\gd_n)$ when $C\qw n\le k\le
Cn$, for each constant $C$.
\end{rem}

The behaviour of the election algorithm is given by the recursion
$X_1=0$ and
\begin{equation}\label{xn}
  X_n\eqd X_{Y_n}+1, \qquad \nn,
\end{equation}
where we assume that $(X_i)_{i=1}^n$ and $Y_n$ are independent.
We state a general convergence theorem for leader election
algorithms of this type.

We recall the definitions of 
the total variation distance $\dtv$ 
and
the Wasserstein distance $\dw$ 
(also known as the Dudley, Fortet-Mourier or Kantorovich distance, or
minimal $L_1$ distance); these are
both metrics on spaces of probability distributions, but it is
convenient to write also $\dtv(X,Y)\=\dtv(\mu,\nu)$ and
$\dw(X,Y)\=\dw(\mu,\nu)$ for random variables $X,Y$
with
$X\sim\mu$ and $Y\sim\nu$.

The total variation distance $\dtv$ between 
(the distributions of) two arbitrary random variables $X$ and $Y$ is
defined by  
\begin{equation}
  \dtv(X,Y)\=\sup_A|\P(X\in A)-\P(Y\in A)|.
\end{equation}
For integer-valued random variables, as is the case in our theorem,
this is easily seen to be equivalent to
\begin{equation}
  \dtv(X,Y)=\tfrac12\sum_k|\P(X=k)-\P(Y=k)|.
\end{equation}
Further, for any distributions $\mu$ and $\nu$,
\begin{equation}\label{dtvcoup}
\dtv(\mu,\nu):=
\inf\left\{\P(X\neq Y):X\sim\mu,\,Y\sim\nu\right\};
\end{equation}
the infimum is taken
over all random vectors $(X,Y)$ on a joint
probability space with the given marginal
distributions $\mu$ and $\nu$.  (In other words,
over all couplings $(X,Y)$ of $\mu$ and $\nu$.) 
For integer-valued random variables,
convergence in $\dtv$
is equivalent to convergence in distribution,
or equivalently, weak convergence of the corresponding distributions.

The Wasserstein distance $\dw$ is defined only for
probability distributions with
finite expectation, and can be defined by, in analogy with \eqref{dtvcoup},
\begin{equation}\label{dwcoup}
\dw(\mu,\nu):=\inf\left\{\E|X-Y|:X\sim\mu,\, Y\sim\nu\right\}.
\end{equation}
There are several equivalent formulas. For example,
for integer-valued random variables, 
\begin{equation}\label{dwinteger}
  \dtv(X,Y)=\sum_k|\P(X\le k)-\P(Y\le k)|.
\end{equation}
It is immediate from \eqref{dtvcoup} and \eqref{dwcoup} that for
integer-valued random variables $X$ and $Y$ (but not in general),
\begin{equation}
  \dtv(X,Y)\le\dw(X,Y).
\end{equation}
It is well-known that 
$\dw$ is a complete metric on the space
of probability measures on $\bbR$ with
finite expectation, and 
that convergence in $\dw$
is equivalent to weak convergence plus convergence
of the first absolute moment.

All unspecified limits in this paper are as $n\to\II$.

\bth       \label{Th1}
Consider the leader election algorithm described above, with 
$Y_n$
satisfying
\refCC. Then,
there exists a distribution function $F$ 
with bounded density function $f=F'$
such that
\begin{equation}
  \label{t1a}
\sup_{k\in\bbZ} |\Pr(X_n\leq k)-F(k-\loga n)|\ra 0
\end{equation}
or, equivalently, if
$Z\sim F$,
\begin{equation}
  \label{t1b}
\dtv(X_n, \lceil Z+\loga n\rceil)\ra 0.
\end{equation}
More precisely,
$
\dw(X_n, \lceil Z+\loga n\rceil)\ra 0,
$
which is equivalent to
\begin{equation}
  \label{t1dw}
\sum_{k\in\bbZ} |\Pr(X_n\leq k)-F(k-\loga n)|\ra 0.
\end{equation}

As a consequence, defining $\DF(x)\=F(x)-F(x-1)$, 
\begin{equation}
  \label{t1adelta}
\sup_{k\in\bbZ} |\Pr(X_n= k)-\DF(k-\loga n)|\ra 0
\end{equation}

Furthermore,
\begin{equation}\label{EXn}
  \E X_n=\loga n+\phi(n)+\bo(1),
\end{equation}
for a continuous function $\phi(t)$ on $(0,\II)$ which is periodic in
$\loga t$, \ie{} $\phi(t)=\phi(\al t)$, and locally Lipschitz.
\ethGL

\begin{figure}[ht]
	\center
		\includegraphics[width=0.55\textwidth,angle=270]{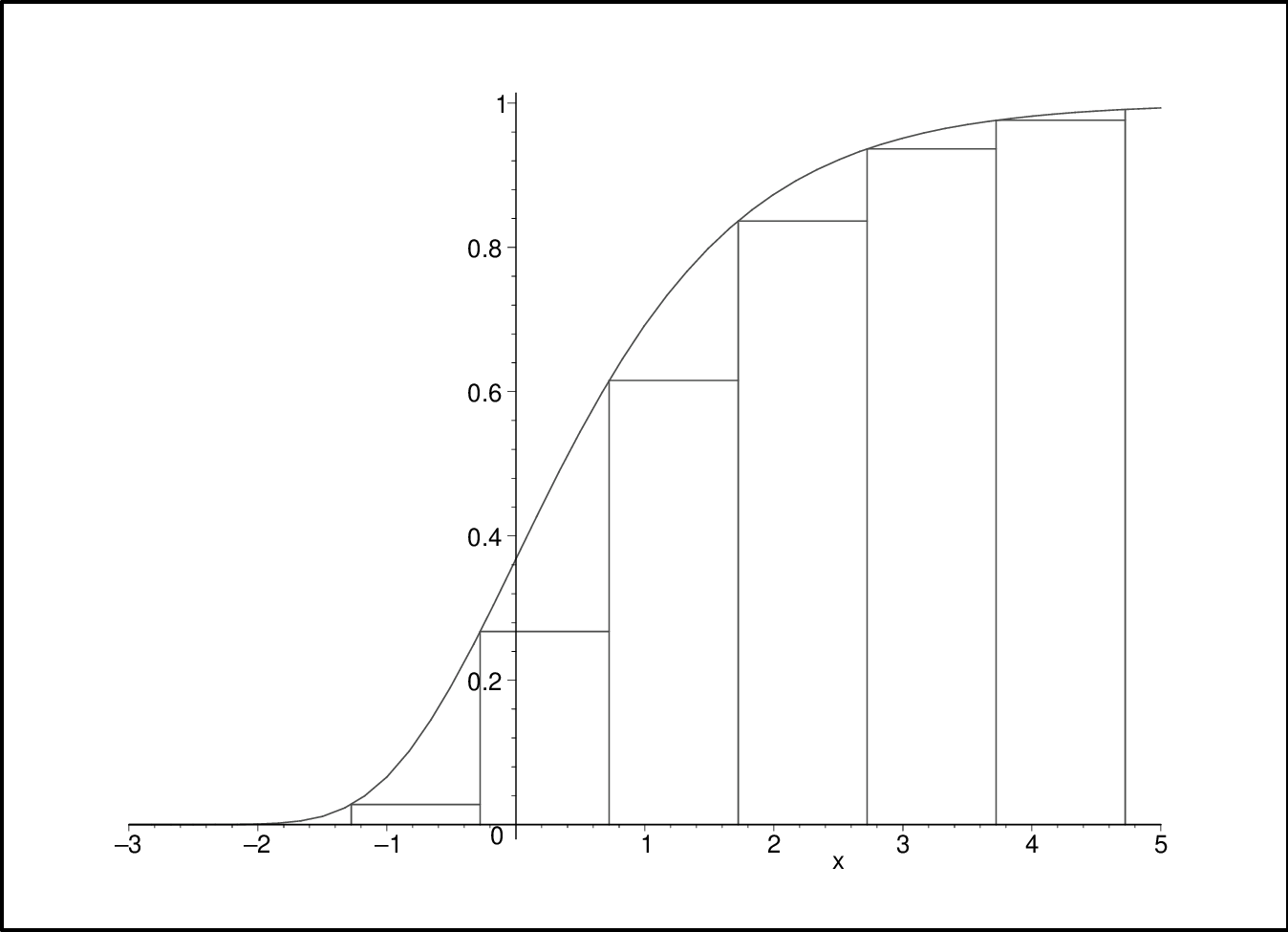}
		\medskip
	\caption{Illustration of \refT{Th1}}
	\label{FP1}
\end{figure}

We thus do not have convergence in distribution as $\ntoo$, but 
the usual type of oscillations with an asymptotic periodicity in $\loga n$
and convergence in distribution along subsequences such that the
fractional part $\set{\loga n}$ converges. (This phenomenon is well-known for
many other problems with integer-valued random variables, 
see for example \cite{LP04,JA04};
it happens frequently when the variance stays bounded.)
This is illustrated in Figure \ref{FP1}.
 
\bpr
We assume that $\gd_n$ are as in \refR{Rdelta}.

Let $q\=\sup_\nn \Pr(Y_n=n)$. 
Since each  $\Pr(Y_n=n)<1$, 
and $\Pr(Y_n=n)\to 0$ by \ref{C1c}, $q<1$. 
Hence $X_n$ is stochastically dominated by a sum of $n-1$ geometric
$\Ge(1-q)$ random variables, and thus $\E X_n=\BO(n)$. 
In particular, $\E X_n<\II$ for every $n$. 

Since the sequence $(Y_n)$ is stochastically increasing, we may couple
all $Y_n$ such that $Y_1\le Y_2\le\dots$. 
If we consider starting our algorithm with different initial values,
and use this coupling of $(Y_n)$ in each round, we obtain a coupling
of all $X_n$, $n\ge1$, such that $X_{n+1}\geq X_n$ \as{} for every $n\ge1$.
We use these couplings of $(Y_n)$ and $(X_n)$ throughout the proof.

Let
\begin{align*}
x_n&:=\E X_n,\\
d_n&:=\E X_{n+1}-\E X_{n}=x_{n+1}-x_n,\\
b_n&:=\max_{1\leq k\leq n} k d_k.
\end{align*}
We extend $b_n$ to real arguments by the same formula; thus,
$b_t=b_{\floor t}$ for real $t\ge1$.

By \eqref{xn},
\[x_n=\E X_n=1+\E x_{Y_n},
\qquad\nn.\]
Thus, for $\nn$,
\beq\label{EJ1}
\begin{split}
d_n
&=\E(x_{Y_{n+1}}-x_{Y_n})
=\E \sum_{Y_n}^{Y_{n+1}-1}d_j
=\E\sum_{j=1}^n d_j \ett[Y_n\leq j < Y_{n+1}]
\\&
=\sum_{j=1}^n d_j \Pr(Y_n\leq j < Y_{n+1}).
\end{split}
\eeq
By \ref{C1b}, $\E Y_{n+1}-\E Y_n\to\a$, and thus there exists $n_0$
such that if $n\ge n_0$ then 
\begin{equation*}
 \sum_{j=1}^n \Pr(Y_n\leq j<Y_{n+1})= \E Y_{n+1}-\E Y_n < 1.
\end{equation*}
Hence \eqref{EJ1} implies, with $d_n^*=\max_{k\leq n} d_k$, 
for $n\ge n_0$,
\begin{equation*}
d_n(1-\Pr(Y_n\leq n<Y_{n+1}))
\leq \sum_{j=1}^{n-1} \Pr(Y_n\leq j<Y_{n+1})d_{n-1}^*
\leq d_{n-1}^* (1-\Pr(Y_n\leq n<Y_{n+1})),
\end{equation*}
and thus
$d_n\leq d_{n-1}^*$ so $d_{n}^*=d_n \vee d_{n-1}^*=d_{n-1}^*$.
Consequently, 
$d_{n}^*=d_{n_0}^*<\II$, for all $n\geq n_0$.
In other words, 
$d^*:=\sup_n d_n <\II$.

Let $\be\=(1+\al)/2$; thus $\a<\gb<1$. 
If $n$ is large enough, so that
$(\al+\de_{n+1})(n+1)<\be n $, then (\ref{EJ1}) yields, using \eqref{c1c} and
\eqref{c1b},  
\bals
d_n
&\leq 
d^* \sum_{j<(\al-\de_n)n}\Pr(Y_n\leq j)
  +\frac{1}{n(\al-\de_n)}b_{\be n}\sum_j \Pr(Y_n\leq j <Y_{n+1})
  +d^*\sum_{j>\be n}\Pr(Y_{n+1}>j)\\ 
&\leq d^* \BO(nn^{-2-\eps})
  +\frac{1}{n(\al-\de_n)}b_{\be n}(\E Y_{n+1}-\E Y_n)\\
&= \BO(n^{-1-\eps})+\frac1n \frac{\al+\BO(\de_n)}{\al-\de_n}b_{\be n}.
\end{align*}
Thus
\beq
n d_n\leq (1+\BO(\de_n))b_{\be n}+\BO(n^{-\eps}).      
\eeq
Replace $n$ by $k$ and take the supremum over all $k$ such that 
$\be n<k\leq n$. Since $b_k$ is increasing, and by our simplifying
assumptions in \refR{Rdelta}, this yields
\[b_n\leq
(1+\BO(\de_{n}))b_{\be n}+\BO(n^{-\eps})
=\bigpar{1+\BO(\de_{n})}b_{\be n}
.\] 
It follows by induction over $m$ that if
$(1/\be)^m \leq n<(1/\be)^{m+1}$,
then
\[b_n\leq \CC \prod_{j=1}^m \Bigpar{1+\frac{\CC}{j^{1+\eps}}}\]
and thus
$b_n=\BO(1)$.
In other words, we have shown
\begin{equation}\label{dn}
  d_n=\BO(1/n).
\end{equation}

We now use the Wasserstein distance $\dw$.
Since $X_{n+1}\geq X_n$ \as, it is easily seen by \eqref{dwcoup} that 
$\dw(X_n,X_{n+1})=\E(X_{n+1}-X_n)=d_n$.
  Thus, if $m\le n$, by \eqref{dn},
\begin{equation*}
  \dw(X_n,X_{m})
\leq \sum_{k=m}^{n-1}   \dw(X_k,X_{k+1})
=
\sum_{k=m}^{n-1} d_k
=
\BO\lp\frac{n-m}{ m}\rp,
\end{equation*}
and thus, for all $n$ and $m$,
  \begin{equation}   \label{E7}
  \dw(X_n,X_{m})
=\BO\lp\frac{|n-m|}{n\wedge m}\rp.
  \end{equation}
Note also that \ref{C1c} implies
\beq
\E|Y_n-\al n|\leq \de_n n+\BO(n^{-1-\eps})=\BO(n\de_n). \label{E4}
\eeq

Define
\begin{equation}
  \label{tX}
\tX_t\=X_{\lfloor t\rfloor}-\loga t,\qquad t\geq 1.
\end{equation}
Then, for $t\geq 2/\al$, using \eqref{xn}, \eqref{E7}, \eqref{E4},
\eqref{c1c},
and $1\le Y_{\lfloor t\rfloor}\le t$,
\begin{equation*}
\begin{split}
\dw(\tX_t,\tX_{\al t})
&
=\dw\bigpar{X_{\lfloor t\rfloor}-\loga(t),X_{\lfloor \al t\rfloor}-\loga(\a t)}
\\&
=\dw\bigpar{X_{\lfloor t\rfloor}-1,X_{\lfloor \al t\rfloor}}
\leq \E \dw\bigpar{X_{Y_{\lfloor t\rfloor}},X_{\lfloor \al t\rfloor}}
\\ &
\le 
\CC\E\lp\frac{|Y_{\lfloor t\rfloor}-\lfloor \al t\rfloor|}
   {Y_{\lfloor t\rfloor}\wedge \lfloor \al t\rfloor}\rp  \\
 &\le 
\CCx\E\lp\frac{|Y_{\lfloor t\rfloor}-\lfloor \al t\rfloor|}
   {\al t/2}+t\ett\bigsqpar{Y_{\lfloor t\rfloor}<{\al t}/{2}}\rp  \\
&= 
\BO\lp t\qw \E|Y_{\lfloor t\rfloor}-\lfloor \al t\rfloor|\rp
+\BO\lp t\Pr\lp Y_{\lfloor t\rfloor}<{\al t}/{2}\rp \rp
\\
  &=\BO(\de_{ \lfloor t\rfloor})+\BO(t\cdot t^{-2-\eps})
=\BO\lp{\log^{-1-\eps} t}\rp.
\end{split}
\end{equation*}
Hence,   
for any $t\geq 2/\al$,
\begin{equation}
\label{dw2}
\sum_{j=0}^\II \dw(\tX_{\al^{-j}t},\tX_{\al^{-j-1}t})
=\BO\lp{\log^{-\eps} t}\rp
<\II.  
\end{equation}
Since $\dw$ is a complete metric, thus  there exists for every $t>0$
a limiting distribution $\mu(t)$, such that if
$Z(t)\sim\mu(t)$, then
\begin{equation}
  \label{XtoZW}
\dw\bigpar{\tX_{\al^{-j}t},Z(t)}\to0
\quad\mbox{as}\quad j\ra \II.
\end{equation}
In particular,
\begin{equation}
\tX_{\al^{-j}t}\dto Z(t) \quad\mbox{as}\quad j\ra \II.
\end{equation}
(We find it more convenient to use the random variable $Z(t)$ than its
  distribution $\mu(t)$.)
  Clearly, $Z(\al t)\stackrel{d}{=} Z(t)$, so
  the distribution $\mu(t)$ is a periodic function of
  $\loga t$. 
Hence, \eqref{XtoZW} can also be written, adding the explicit estimate
  obtained from \eqref{dw2},
\begin{equation}
  \label{XtoZWt}
\dw\bigpar{\tX_{t},Z(t)}
=\BO\lp{\log^{-\eps} t}\rp
\to0
\quad\mbox{as}\quad t\ra \II.
\end{equation}

Note further that, for $\gam\geq 1$, by \eqref{tX} and \eqref{E7},
\begin{equation*}
  \begin{split}
\dw(\tX_t,\tX_{ \gam t})
&\leq \dw(X_{\lfloor  t\rfloor}, X_{\lfloor \gam t\rfloor})
   +|\loga t-\loga ( \gam t)| 
=\BO\lp \frac{\lfloor  \gam t\rfloor-\lfloor t\rfloor}{t}\rp  +\loga  \gam
\\&=\BO( \gam-1+1/t).
  \end{split}
\end{equation*}
Replacing $t$ by $\al^{-j}t$ and letting  $j \ra \II$, it follows
from \eqref{XtoZW} 
that, for all $t>0$ and $\gam\ge1$, 
\begin{equation}
  \label{dwz}
 \dw(Z(t),Z( \gam t))=\BO(\gam-1).
\end{equation}
Consequently, 
$t\ra \mu(t)=\mathcal{L}(Z(t))$ is continuous and Lipschitz in the
Wasserstein metric. 

 Define, for every real $x$,
 \begin{equation}
   \label{F}
F(x)=\Pr (Z(t) \le x)
 \end{equation}
for any $t>0$ such that $x+\loga t$ is an integer;
since $Z(t)$ is periodic in $\loga t$, this does not depend on the
choice of $t$. 

Since $\tX_{\al^{-j}t}+\loga t = X_{\al^{-j}t}-j\in\bbZ$,
the random variable $Z(t)+\loga t$ is integer-valued for every $t$. 
It is easily
seen that for integer-valued random variables $Z_1$ and $Z_2$, the
total variation distance $\dtv(Z_1,Z_2)\le\dw(Z_1,Z_2)$.
Hence, for any $x\in\bbR$ and $u\ge0$, choosing $t$ such that 
$x+\loga t$ is an integer and letting $\gam=\al^{-u}$, which implies
that $x-u+\loga(\gam t)=x+\loga(t)\in\bbZ$, we obtain from the
definition \eqref{F} and \eqref{dwz},
\begin{equation}
  \begin{split}
F(x)-F(x-u)
&=
\P\bigpar{Z(t)\le x}  - \P\bigpar{Z(\gam t)\le x-u}	
\\&
=
\P\bigpar{Z(t)+\loga t\le x+\loga t}  
- \P\bigpar{Z(\gam t)+\loga(\gam t)\le x+\loga t}	
\\&
\le \dtv\bigpar{Z(t)+\loga t,\,Z(\gam t)+\loga(\gam t)}
\\&
\le \dw\bigpar{Z(t)+\loga t,\,Z(\gam t)+\loga(\gam t)}
\\&
\le \dw\bigpar{Z(t),Z(\gam t)}+|\loga t-\loga(\gam t)|
\\&
=\BO(\gam -1)+\loga \gam 
=\BO(u).
  \end{split}
\raisetag{\baselineskip}
\label{dF}
\end{equation}
Hence, $F(x)$ is a continuous function of $x$.

We have shown that 
$t\ra \mathcal{L}(Z(t))$ is continuous in the
Wasserstein metric, and thus in the usual topology of weak
convergence in the space $\cP(\bbR)$ of probability measures on
$\bbR$. Since further $\cL(Z(t))$ is 
periodic in $t$, the set 
$\set{\cL(Z(t)):t>0}=\set{\cL(Z(t)):1\le t\le\al\qw}$ is compact in
$\cP(\bbR)$, which by Prohorov's theorem means that the family $\set{Z(t)}$
of random variables is tight, see \eg{} Billingsley \cite{BI68}.
Hence, $\P(Z(t)\le x)\to0$ as $x\to-\infty$ and 
$\P(Z(t)\le x)\to1$ as $x\to+\infty$, uniformly in $t$, and it follows
from \eqref{F} that
$\lim_{x\to-\infty} F(x)=0$ and
$\lim_{x\to\infty} F(x)=1$.

Furthermore, \eqref{XtoZWt} and \eqref{F} show that, for any sequence $k_n$ of
integers, as $n\to\II$, 
\begin{equation}\label{pxkn}
\begin{split}
\Pr(X_n \le k_n)
&=
\Pr\bigpar{\tX_n \le k_n-\loga n}
=
\Pr\bigpar{Z(n) \le k_n-\loga n} + \bo(1)
\\&
=
F( k_n-\loga n)+\bo(1).
\end{split}
\end{equation}
Since further the sequence $X_n$ is increasing, 
it now follows from Janson \cite[Lemma 4.6]{JA04}
that $F$ is monotone, and thus a distribution function.
By \eqref{dF}, the distribution is absolutely continuous and has a
bounded density function $F'(x)$.

It is easy to see that \eqref{pxkn}, \eqref{t1a} and \eqref{t1b} are
equivalent, see \cite[Lemma 4.1]{JA04}. The corresponding result in
the Wasserstein distance follows from \eqref{XtoZWt} because
$\dw(X_n, \lceil Z+\loga n\rceil)
=\dw(\tX_n, Z(n))$, \eg{} by \refR{RZ} below;
\eqref{t1dw} then follows by \eqref{dwinteger}.
Finally, \eqref{XtoZWt} implies that
\[
\E \tX_t=\E X_{\lfloor t\rfloor}-\loga t=\E Z(t)+\BO\lp\log^{-\eps} t\rp,
\]
which proves \eqref{EXn} with  $\phi(t)\=\E Z(t)$, which is periodic
in $\loga t$. 
Since $|\phi(t)-\phi(u)|=|\E(Z(t)-Z(u))|\le\dw(Z(t),Z(u))$, \eqref{dwz}
implies that $\phi$ is continuous, and Lipschitz on compact intervals.
\epr

\begin{rem}\label{RZ}
  As remarked above, $Z(t)+\loga t$ is integer-valued. Moreover, for
  every integer $k$,
  \begin{equation*}
	\begin{split}
\P\bigpar{Z(t)+\loga t\le k}	  
&=\P\bigpar{Z(t)\le k-\loga t}
=F(k-\loga t)
\\&
=\P\bigpar{Z\le k-\loga t}
=\P\bigpar{Z+\loga t\le k}
\\&
=\P\bigpar{\ceil{Z+\loga t}\le k}.
	\end{split}
  \end{equation*}
Hence, for every $t>0$, 
$Z(t)\eqd \lceil Z+\loga t\rceil -\loga t$.
General families of random variables of this type are studied 
in \cite{JA04}.
In particular, \cite[Theorem 2.3]{JA04} shows how $\phi(t)\=\E Z(t)$
in \refT{Th1}
can be obtained from the characteristic function of the distribution
$F$ of $Z$.
\end{rem}

\begin{rem}
The very slow convergence rate 
$\BO\lp{\log^{-\eps} t}\rp$
in \eqref{XtoZWt} is because we allow $\gd_n$ to tend to 0 slowly.
In typical applications, $\gd_n=n^{-a}$ for some $a>0$, and then
better convergence rates can be obtained. We have, however, 
not pursued this.
\end{rem}

\begin{rem}
  Note that $F$ and $\phi$ are influenced by the distribution of $Y_n$
  for small $n>2$, for example $Y_3$ and $Y_4$; hence there is no hope for a
  nice explicit formula for $F$ or $\phi$ depending only on asymptotic
  properties of $Y_n$. 
\end{rem}

\begin{rem}
The general problem of studying the number of steps until absorption
at 1 of a decreasing Markov chain on \set{1,2,\dots} appears in many
other situations too, usually with quite different behaviour of $Y_n$
and $X_n$. As examples we mention the recent papers studying random
trees and coalescents by
Drmota, Iksanov, Moehle and Roesler \cite{DIMR06},
Iksanov and  M\"ohle \cite{IkM07}
and
Gnedin and Yakubovich \cite{GneY06}; in these papers the number killed
in each round is much smaller than here and thus $X_n$ is larger, of
the order $n$ or $n/\log n$; moreover, after normalization $X_n$ has a
stable limit law. 
\end{rem}

\section{Extensions}\label{Svar}

We have assumed that we repeat the elimination step until only one
player remains. As a generalization we may suppose that we stop when
there are at most $a$ players left, for some given number $a$.

\bth       \label{Th1a}
Consider the leader election algorithm described in \refS{SJ0}, 
but stopping as soon as the number of remaining
players is at most $a$, for some
fixed $a\ge1$.
Suppose that \refCC{} is satisfied. Then, the conclusions of
\refT{Th1} hold, for some
$F$ and $\phi$ that depend on the threshold $a$.
\ethGL

\begin{proof}
This generalization can be obtained from the version in \refS{SJ0} by
replacing $Y_n$ by
\begin{equation*}
  Y_n'\=
  \begin{cases}
Y_n, & Y_n>a;
\\
1,& Y_n\le a.	
  \end{cases}
\end{equation*}

Suppose that \refCC{} holds for $(Y_n)$.
It is easily seen that then \refCC{} holds for $(Y_n')$ too, with the
same $\al$; for \ref{C1b}, note that \refCC\ref{C1c} implies that
\begin{equation*}
  \E|Y_n-Y_n'|\le a\P(Y_n\le a) = \BO(n^{-2-\eps})
\end{equation*}
and thus $\E Y_{n+1}'-\E Y_n' =  \E Y_{n+1} - \E Y_n + \bo(n^{-2})$.
Consequently, \refT{Th1} applies to $(Y_n')$, and the result follows.
\end{proof}

In this situation, it is also interesting to study the probability
$\pi_i(n)$ that the procedure ends with exactly $i$ players, starting
with $n$ players; here $i=1,\dots,a$ and $\sum_{i=1}^a\pi_i(n)=1$.
We have a corresponding limit theorem for $\pi_i(n)$.
\bth       \label{Thpi}
Suppose that \refCC{} holds and that $a\ge1$ is given as in \refT{Th1a}. Then,
\begin{equation}\label{pin}
  \pi_i(n)=\psi_i(n)+\bo(1),
\qquad i=1,\dots,a,
\end{equation}
for some
continuous functions $\psi_i(t)$  on $(0,\II)$ which are periodic in
$\loga t$, \ie{} $\psi_i(t)=\psi_i(\al t)$, and locally Lipschitz.
\ethGL

\begin{proof}
A modification of the proof of \refT{Th1}, now taking $x_n\=\pi_i(n)$
and $d_n\=|x_{n+1}-x_n|$ and replacing the random $\tX_t$ by
$x_{\floor t} =\pi_i(\floor t)$, 
yields $\pi_i(n+1)-\pi_i(n)=\BO(1/n)$ and 
$\pi_i(\al^{-j}t) - \pi_i(\al^{-(j+1)}t) = \BO(j^{-1-\eps})$;
hence, for any $t>0$, $\pi_i(\al^{-j}t)\to\psi_i(t)$ for some
$\psi_i(t)$, which easily is seen to satisfy the stated conditions.
We omit the details.
\end{proof}

More generally, there is a similar result on the probability that the
process passes through a certain state; this is interesting also for
the process in \refS{SJ0} with $a=1$.

\bth       \label{Thpi2}
Suppose that \refCC{} holds and that $a\ge1$ is given as above.
Let $\pi_i(n)$, $i\ge1$, be the probability that, starting with $n$
players, there exists some round with exactly $i$ survivors.
Then
\begin{equation}\label{pin2}
  \pi_i(n)=\psi_i(n)+\bo(1),
\qquad i=1,2,\dots,
\end{equation}
for some
continuous functions $\psi_i(t)$  on $(0,\II)$ which are periodic in
$\loga t$, \ie{} $\psi_i(t)=\psi_i(\al t)$, and locally Lipschitz.
\ethGL

\begin{proof}
For $i\le a$, this $\pi_i(n)$ is the same as in \refT{Thpi}, and
for each $i> a$, this $\pi_i(n)$ is the same as in \refT{Thpi} if we
replace $a$ by $i$.
\end{proof}

\begin{rem}\label{R0}
Another variation, which is natural in some problems, is to study a
non-increasing Markov chain on \set{0,1,\dots} and ask for the number
of steps to reach 0; in other words, the time until all players are
killed. In this case, we thus assume that $0\le Y_n\le n$. This can
obviously be transformed to our set-up on \set{1,2,\dots} by
by increasing each integer by 1; in other words, we replace $Y_n$ by
$Y_n'\=Y_{n-1}+1$, $n\ge2$; we can interpret this as adding a dummy
player that never is eliminated, and continuing until only the dummy remains.
If \refCC{} holds for $Y_n$, except that $Y_n=0$ is allowed and
$\P(Y_1=0)>0$, then 
\refCC{} holds for $Y_n'$ too, and thus our results hold also in this
case, with $X_n$ now defined as the number of steps until
absorption in 0. (To be precise, $X_n=X'_{n+1}$, with $(X'_n)$
corresponding to $(Y_n)$, since we add a dummy, but 
there is no difference between the asymptotics of $X_{n+1}'$ and $X_n'$.)
\end{rem}

\section{Examples}\label{Sex}

\begin{example}[a toy example] \label{Ex0}
  For a simple example to illustrate the theorems above, let, for
  $n\ge 2$,
$Y_n=\floor{(n+I)/2}$, where $I\sim\Be(1/2)$ is 0 or 1 with
  $\P(I=1)=1/2$. In other words, we toss a coin and let $Y_n$ be
  either $\floor{n/2}$ or $\ceil{n/2}$ depending on the outcome. (If
  $n$ is even, thus always $Y_n=n/2$.)
Note that $\E Y_n=n/2$, $n\ge2$, and that \refCC{} holds trivially,
  with $\al=1/2$.
If we start with $N_0=n$ players and $m 2^j \le n\le (m+1)2^j$,
  $m\ge1$, then the number $N_j$ of survivors after $j$ rounds
  satisfies $m\le N_j\le m+1$ and $\E N_j = 2^{-j}n$ (by induction on
  $j$).
Consequently,
if $m 2^j \le n\le (m+1)2^j$,
\begin{align}
 \label{e0}
\P(N_j=m)=m+1-2^{-j}n,
&&&
\P(N_j=m+1)=2^{-j}n-m.
\end{align}
Taking $m=1$, this shows that if $2^j\le n\le 2^{j+1}$, then
\begin{align}
\P(X_n=j)=\P(N_j=1)=2-2^{-j}n,
&&&
\P(X_n=j+1)=1-\P(X_n=j)=2^{-j}n-1.
\end{align}
Hence, \eqref{t1a} holds exactly,
$\P(X_n\le k)=F(k-\log_2n)$, for all $k\in\bbZ$ and $n\ge1$,
with
\begin{equation*}
  F(x)=
  \begin{cases}
0, & x\le-1,
\\
2-2^{-x},& -1\le x\le0,
\\
1,& x\ge0,	
  \end{cases}
\end{equation*}
and \eqref{EXn} holds exactly, $\E X_n=\log_2 n+\phi(n)$, with
\begin{equation*}
  \phi(2^x)=2^{x-\floor x} - \bigpar{x-\floor x}-1.
\end{equation*}

Suppose now, as in \refS{Svar}, that we stop when there are at most $a=3$
players left. (Similar results are easily obtained for
other values of $a$.)
If $2^i\le n\le2^{i+1}$ with $i\ge1$, we take $j=i-1$ 
and note that
$N_j\in\set{2,3,4}$.
If $N_j=4$, the procedure ends after one further round with 2 players
left; otherwise it ends immediately with $N_j=2$ or 3 survivors.
Taking $m=2$ or $m=3$ in \eqref{e0}, we thus find
\begin{equation*}
\pi_2(n)=
\P(N_{i-1}=2)+\P(N_{i-1}=4)
=|2^{1-i}n-3|,
\qquad 
2^i\le n\le2^{i+1}.
\end{equation*}
Consequently, $\pi_2(n)=\psi_2(n)$ exactly, for all $n\ge2$, with
\begin{equation*}
\psi_2(2^x)=
|2^{1+x-\floor x}-3|.
\end{equation*}
Further, $\psi_3(t)=1-\psi_2(t)$ and $\psi_1(t)=0$.
\end{example}

\begin{example}[a counter example]
The procedure in \refE{Ex0} is almost deterministic. In contrast, the
very similar but 
completely deterministic $Y_n=\floor{n/2}$, $n\ge2$, does not
satisfy \refCC\ref{C1b}. In this case, $X_n=\floor{\log_2 n}$ and 
$\P(X_n\le k)=F(k-\log_2n)$, for all $k\in\bbZ$ and $n\ge1$,
where $F(x)=\ett[x\ge0]$, the distribution of $Z\=0$; this limit $F$ is
not continuous so the conclusions of \refT{Th1} do not all hold.
\end{example}

\begin{example}\label{Efair}
A leader election algorithm studied by Prodinger \cite{PR93},
Fill,  Mahmoud and Szpankowski \cite{FMS96}
Knessl \cite{Kn01}, 
and Louchard and Prodinger 	\cite{LPasymmetric},
see also Szpankowski \cite[Section 10.5.1]{SZ01},
is the following:
\emph{%
Each player tosses a fair coin. If at least one player throws heads,
then all players throwing tails are eliminated; if all players throw
tails, then all survive until the next round.
}

Except for the special rule when all throw tails, which 
guarantees that at least one player survives each round,
the number $Y_n$ of survivors in a round thus has a binomial
distribution $\Bi(n,\tfrac12)$.
More precisely, if $W_n$ is the number of heads thrown,
\begin{equation}\label{tails}
Y_n=W_n+n\ett[W_n=0]
\qquad\text{with }  
W_n\sim\Bi(n,\tfrac12).
\end{equation}

Note that 
\begin{equation*}
  \E|Y_n-n/2|^6
=
  \E|W_n-n/2|^6
=\BO(n^3),
\end{equation*}
so \eqref{ymom} holds for $p=6$ (and, indeed, for any $p>0$), and thus
\eqref{c1c} holds. Similarly,
\begin{equation*}
  \E Y_n=\E W_n+2^{-n} n = \tfrac12 n + n 2^{-n}.
\end{equation*}
Thus, conditions \ref{C1b} and \ref{C1c} in \refT{Th1} are
satisfied. Also the monotonicity condition \ref{C1a} is satisfied,
because if $1\le k\le n-1$ (other cases are trivial), then
\begin{equation}\label{c1afair}
  \begin{split}
  \P(Y_{n+1}\le k)
&=
  \P(1\le W_{n+1}\le k)
= \tfrac12 \P(1\le W_{n}\le k) +  \tfrac12 \P(0\le W_{n}\le k-1)
\\&
=
 \P(Y_{n}\le k) +  \tfrac12 \bigpar{\P(W_{n}=0) - \P(W_{n}= k)}
\\&
< \P(Y_{n}\le k).	
  \end{split}
\end{equation}
Hence \refCC{} is satisfied with $\al=1/2$ and \refT{Th1} applies.

In this case, 
Prodinger \cite{PR93}, see also 
Fill,  Mahmoud and Szpankowski \cite{FMS96},
found an exact formula for the expectation 
$\E X_n$ and asymptotics of the form \eqref{EXn} with the explicit
function
\begin{equation}\label{dlw3}
\phi(t)
=\tfrac12-(\log 2)\qw\sum_{k\neq 0}\zeta(1-\chi_k)\Gamma(1-\chi_k)
e^{2k\pi\ii\log_2 t},
\qquad
\chi_k\=\frac{2k\pi}{\log 2}\ii.
\end{equation}
Fill,  Mahmoud and Szpankowski \cite{FMS96} further found asymptotics
of the distribution that can be written as \eqref{t1a} with
\begin{equation}
  F(x)=\frac{2^{-x}}{\exp(2^{-x})-1},
\end{equation}
which thus is the distribution function of $Z$ in this case.
Second and higher moments are considered by Louchard and Prodinger
\cite{LPasymmetric}. 

Prodinger \cite{PR93} considered also the possibility of stopping at
$a=2$ players, and showed \eqref{EXn} above in this case too,
with an explicit formula
for the function $\phi(t)$ (of the same type as \eqref{dlw3} for $a=1$).
\end{example}

\begin{example}\label{Ebias}
A variation of \refE{Efair} 
studied by Janson and Szpankowski \cite{JS97},
Knessl \cite{Kn01} 
and Louchard and Prodinger 	\cite{LPasymmetric}
is to let the coin be biased, with probability
$p\in(0,1)$ for heads (=survival). 
Then \eqref{tails} still holds, but with $W_n\sim\Bi(n,p)$.
Conditions \ref{C1}\ref{C1b}\ref{C1c} hold as above, 
with $\al=p$,
but arguing as in
\eqref{c1afair} we see that \refCC\ref{C1a} holds if $p\ge 1/2$, but
not for smaller $p$. Hence, \refT{Th1} applies when $p\ge1/2$.

In fact, the results of 
\cite{JS97}
show that the conclusions
\eqref{t1a} and \eqref{EXn} hold for all $p\in(0,1)$, for some
functions $F$ and $\phi$ explicitly given in 
\cite{JS97}. 
(See also 
\cite{LPasymmetric}, where further higher moments are treated.)
This suggests that \refT{Th1} should hold more
generally. Note that although \refCC\ref{C1a} does not hold for
$p<1/2$, the difference 
$\P(Y_{n+1}\le k)-\P(Y_{n}\le k)$ is at most $\P(Y_n=0)=(1-p)^n$,
and thus negative or exponentially small for large $n$. It seems likely that
\refT{Th1} can be extended to such cases, by allowing a small
error in \refCC\ref{C1a}; this then would
include this leader
election algorithm with a bisaed coin for any $p\in(0,1)$. However, we
have not pursued this.

It was left as an open question in \cite{JS97} whether 
for each $p\in(0,1)$
the limit
function $F$ is monotone, and thus a distribution function, which
means that there exists a random variable $Z$ such that \eqref{t1b}
holds. By the discussion above, \refT{Th1} shows that this holds for
$p\ge1/2$, but the case $p<1/2$ is as far as we know still open.
Cf.\ \refR{Rnonmonotone} and Figure \ref{F5} below, which show that
monotonicity fails in a related situation. 
Numerical experiments, based on \cite[Prop.\ 3.1]{LPasymmetric}, 
indicate that $F$ is  monotone, at least for some choices of $p<1/2$. 
\end{example}

A further variation of \refE{Efair} is to let the probability $p$ depend on
$n$. The case $p=1/n$ is studied by Lavault and Louchard
\cite{LL06}; in this case $\E Y_n$ is bounded and \refCC{} does not
hold, so \refT{Th1} does not apply.

\begin{example}\label{Exit}
The special rule in Examples \refand{Efair}{Ebias} 
for the exceptional case when all throw tails is of course necessary
to prevent us from killing all players, but as we have seen, it
complicates the analysis, especially for $p<1/2$ when it destroys stochastic
monotonicity of $Y_n$. Note that this rule typically is invoked only
towards the end of the algorithm, when only a few players are left. We
regard the rule as an emergency exit, and it could be replaced by other
special rules for this case. For example, an alternative would be to
switch to some other algorithm that is fail-safe although in principle
(for large $n$) slower; for our purpose this means that the present
algorithm terminates, so we may describe this by letting $Y_n=1$ in
this case, \ie, \eqref{tails} is replaced by $Y_n=W_n+\ett[W_n=0]=\max(W_n,1)$.
Note that for this version, \refCC{} holds for every $p\in(0,1)$, with
$\al=p$, so our theorems apply.

An equivalent way to treat this version is to add (as in \refR{R0})
a dummy, which is
exempt from elimination, and to eliminate everyone else that throws a
tail; we then stop when there are at most 2 players left (the dummy and,
possibly, one real player). This is thus the version in \refS{Svar},
with $a=2$ and $Y_n=1+W_{n-1}$, $W_m\sim\Bi(m,p)$ (and starting with
$n+1$ players). Again, \refCC{} holds, with $\al=p$, and the results
in \refS{Svar} apply. In particular, since invoking the special rule
corresponds to eliminating everyone except the dummy, the probability
that we have to invoke the special rule is the same as the probability
that the dummy version ends with only the dummy, \ie{} $\pi_1(n+1)$ in
the notation of \refT{Thpi}; the asymptotics of this probability is
thus given by \eqref{pin}.
\end{example}

\begin{example}\label{Edemon}
Prodinger \cite{prodinger93} (for $p=1/2$)
and
  Louchard and Prodinger \cite{LPdemon} studied a version of Examples
  \refand{Efair}{Ebias} where, as in \refR{R0}, we allow all players
  to be killed and let $X_n$ be the time until that happens.
Thus, in each round,
each player tosses a coin and is killed with probability $1-p$ (and we
  do not have any special rule). Additionally, there is a demon, who
  in each round kills one of the survivors (if any) with probability
  $\nu\in[0,1]$.
We thus have the modification discussed in \refR{R0}, with
\begin{equation*}
  Y_n=\max(W_n-I,0),
\qquad W_n\sim\Bi(n,p),\; I_\nu\sim\Be(\nu), 
\end{equation*}
where $W_n$ and $I_\nu$ are independent; thus also
$Y_n'=Y_{n-1}+1=\max(W_{n-1}+1-I_\nu,1)$.
\refCC{} holds, in the modification for absorption in
0, and thus our 
results apply. In fact, 
Louchard and Prodinger \cite{LPdemon} show
\eqref{t1a} and \eqref{EXn} for this problem
with explicitly given $F$ and $\phi$; they
further give an extension of \eqref{EXn} to higher moments.

As remarked in \cite{prodinger93,LPdemon}, 
the special case $\nu=1$ is equivalent
to approximate counting
and $\nu=0$ is equivalent to
the cost of an unsuccessful search in a trie; in the latter case, 
$X_n$ is simply the maximum of $n$ i.i.d\ geometric
random variables which can be treated by elementary methods.
\end{example}

\begin{example}\label{Efranklin}
W.R. Franklin \cite{FR82} proposed
a leader election algorithm where
the $n$ players are arranged
in a ring. Each player gets a random number; these are i.i.d.\ and,
say, uniform on $[0,1]$. 
(Since only the order of these numbers will matter, any continuous
distribution will do; moreover, 
it is equivalent to let $\xi_1,\dots,\xi_n$ be a random
permutation of $1,\dots,n$.)
A player survives the first round if her random number is a peak; in
other words, if $\xi_1,\dots,\xi_n$ are i.i.d.\ random numbers, then 
player $i$ survives if $\xi_i\ge\xi_{i-1}$ and $\xi\ge\xi_{i+1}$
(with indices taken modulo $n$). We may ignore the possibility that
two numbers $\xi_i$ and $\xi_j$ are equal; hence we may as well 
require
$\xi_i>\xi_{i-1}$ and $\xi>\xi_{i+1}$.

In Franklin's algorithm, the survivors continue by comparing their
original numbers in the same way with the nearest surviving players;
this is repeated until a single winner remains.
We have so far not been able to analyse this algorithm. It is easy to
verify that even if we condition on the number $m$ of survivors after
the first round, the $m!$ possible different orderings of the
survivors do not appear with equal probabilities, which means that the
algorithm is \emph{not} of the recursive type studied in this paper. 
(For example, starting with a 
ring of 8
players and conditioning on having 4 survivors (peaks) in the first round; the
probability of getting 2 survivors in the second round is 10/34, and
not 1/3 as in the uniform case.) 

However, we can study a variation of Franklin's algorithm, where the
survivors draw new random numbers in each round. This is an algorithm
of the type studied in \refS{SJ0}, with $Y_n$ given by the number of
peaks in a random permutation, regarded as a circular list. Note that
there is always at least one peak (the maximum will always do), so
$Y_n\ge1$ as required. It is easily seen that inserting a new player will
never decrease the number of peaks; hence $Y_n$ is stochastically
increasing in $n$. Further, we have $Y_n=\sum_1^n I_i$, where 
$I_i\=\ett[\xi_i>\max(\xi_{i-1},\xi_{i+1})]$ is the indicator that
player $i$ survives (again, indices are taken modulo $n$). 
If $n\ge3$, then $\E I_i=1/3$ by symmetry, 
and thus $\E Y_n=n/3$. In particular, \ref{C1b} holds with $\alpha=1/3$.
Furthermore, $I_i$ and $I_j$ are independent unless $|i-j|\le2$ $\pmod
n$, and similarly for sets of the indicators $I_i$, and it follows
easily that $\E (Y_n-\E Y_n)^6=\BO(n^3)$, and thus \eqref{ymom} holds
for $p=6$. (Indeed, \eqref{ymom} holds for all even $p$ by this
argument.)
Consequently, \refCC{} holds and \refT{Th1} applies, with $1/\alpha=3$.
\end{example}

\begin{rem}
It might be shown that for the true Franklin algorithm, the expected number of
survivors after 2 rounds is $c_2 n+o(n)$, where
\begin{equation}\label{franklin2}
  c_2=
\frac{3e^4-48e^2+233}{384}
\approx
0.1096868681.
\end{equation}
(Details might  appear elsewhere.)  
In comparison, for the variation with new random numbers each round,
it is easily seen that the expected number after $k$ rounds is
$(1/3)^k n+\bo(1)$, for any fixed $k$; in particular, after two rounds
it is $n/9+\bo(n)$. 
Note that $c_2$ in \eqref{franklin2} is
slightly smaller than $1/9$, and thus better in terms of performance.
It might have been hoped that the true Franklin algorithm is asymptotically
equivalent to the variation studied here, but
the fact that $c_2\neq1/9$ suggests that this is not the case.
Nevertheless, we conjecture that \refT{Th1} remains true for the
Franklin algorithm, for some unknown $\alpha<1/3$.
\end{rem}

\begin{example}   \label{EX39}   
In both versions in \refE{Efranklin}, the players are arranged in a
circle.
Alternatively, the players may be arranged in a line. We use the same
rules as above, but we have to specify when a player at the end (with
only one neighbour) is a peak. There are two obvious possibilities:
\begin{romenumerate}
  \item
Never regard the first and last players as peaks.
(Define $\xi_0=\xi_{n+1}=+\infty$.)
\item
Regard them as peaks if $\xi_1>\xi_2$ and $\xi_n>\xi_{n-1}$,
respectively.
(Define $\xi_0=\xi_{n+1}=-\infty$.)
\end{romenumerate}
In the first case, it is possible that there are no peaks, and thus we
have to add an emergency exit as in \refE{Exit}.

As in \refE{Efranklin}, there are two versions (for each of (i) and
(ii)): we may use the same random numbers in all rounds, or we may
draw new ones each round.

In the latter case, we are again in the situation of \refS{SJ0}.
In both cases (i) and (ii), the distribution of $Y_n$ is related to the
distribution in the circular case in \refE{Efranklin}.
Indeed, if we start with a circular list of $n+1$ numbers and
eliminate the player with the largest number, then the remaining $n$
numbers form a linear list, and the peaks in this list
using version (i) equal the peaks except the maximum one in the
original circular list. 
Similarly, if we instead eliminate the player with the smallest
number, then the peaks in the remaining list using version (ii) equal
the peaks in the original list.
Hence, if $Z_n$ is the number of peaks in a
random circular list of length $n$, then 
(i) yields $Y_n\eqd Z_{n+1}-1$ and 
(ii) yields $Y_n\eqd Z_{n+1}$.
In both cases, this implies that \refCC{} holds (provided we add a
suitable emergency exit in case (i)) because it holds for
$Z_n$. Consequently, \refT{Th1} applies to both these linear versions
of (the variation of) Franklin's algorithm, again with $1/\alpha=3$.
\end{example}

\section{First variation of the Franklin leader election algorithm. The linear case}\label{SS1}
We assume that the survivors draw new random numbers in each round and that they are arranged in line. We use possibility (i) of \refE{EX39}. 
We start with a set of $n$ players. We assign a classical permutation
of $\{1,\dots,n\}$ to the set, all players corresponding to  a peak stay
alive, the other ones are killed. If there are no peaks, we choose the following emergency exit: a
player is chosen at random (this is assumed to have $0$ cost), indeed in the
original game, one deals with circular permutations, so there always
exists at least one peak, here we  approach the problem with a
classical inline permutation. 

What is the distribution of the number $X_n$ of \emph{phases} (or
rounds) before getting only one player? 

\subsection{The analysis}\label{S0}
Let
\bals
Y_n&:=\mbox{ number of peaks, starting with }n\mbox{ players},\\
P(n,k)&:=\Pr[Y_n=k]=\Pr [k\mbox{ peaks, starting with }n\mbox{ players}],\\
\Pi(n,j)&:=\Pr(X_n=j)=\Pr[j \mbox{ phases are necessary to end the
	game, starting with }n\mbox{ players}],\\ 
\La(n,j)&:=\sum_{k=0}^{j}\Pi(n,k)
=
\Pr[\text{at most $j$ phases are necessary, starting with }n\mbox{ players}].
\end{align*}
We will sometimes use the subscript $\xxl$ to distinguish these from
the circular case discussed in \refS{SS2}.

First of  all, we know (Carlitz \cite{Ca74}; see also \cite[Chapter 3]{FS07}), 
that the pentavariate generating function (GF) of valleys ($u_0$),
double rise ($u_1$), double fall ($u'_1$) 
and peaks ($u_2$) is given by
\[I(z,\mathbf{u})=\frac{\de}{u_2}\frac{v_1+\de
  \tan(z\de)}{\de-v_1\tan(z\de)}-\frac{v_1}{u_2},\] 
with
\[v_1=(u_1+u'_1)/2,\quad \de=\sqrt{u_0u_2-v_1^2}.\]
This gives the GF of the number of peaks:
\beq
\frac{\tan[z(u-1)^{1/2}]}{(u-1)^{1/2}-\tan[z(u-1)^{1/2}]} ,
\label{E0} 
\eeq
hence the mean $\M$ and variance $\V$ of the number of peaks,
for $n\ge2$ and $n\ge4$, respectively:
\begin{equation}
  \label{mv}
\M(n)=(n-2)/3,\quad \V(n)=2(n+1)/45.
\end{equation}
This GF is also given in Carlitz \cite{Ca74}.
Moreover, from \cite[Chapter 9]{FS07}, we know that the distribution
$P$ is asymptotically Gaussian. This is also proved in Esseen
\cite{Es82} by probabilistic methods. 

Let $x(n)$ be the mean number of phases, $\E(X_n)$, starting with $n$
players. As we shall see, the initial values are 
\[x(0)=x(1)=0,\quad x(2)=x(3)=x(4)=1.\]
Since \eqref{mv} yields $M(n)+1=(n+1)/3$ for $n\ge2$, we have
(approximating by using this for $n\le1$ too)
that
the mean number of players $c(j)$ still alive after $j$ phases is 
\[c(j)\approx3^{-j}(n+1)-1.\]
(An induction easily yields the exact formula $|c(j)-3^{-j}n|<1$ for
all $j$.)
If we want $c(j)=1$, this leads to the approximation
\[x(n)\approx j\approx \log_3 n-\log_3 2.\]
We see from \refT{Th1} (which applies by \refE{EX39}) that this is
roughly correct, but the constant  $-\log_3 2$ has to be replaced by a
periodic function $\phi(n)$.

Let us now construct $\Pi$. We have,
for $n\ge2$ and $j\ge1$,
\beq
\Pi(n,j)=\sum_{k=0}^{\lfloor (n-1)/2\rfloor}P(n,k)\Pi(k,j-1).
\label{E1} 
\eeq
We have  the initial values
\[\Pi(0,0)=1,\;\Pi(0,j)=0,j>0,\;\Pi(1,0)=1,\;\Pi(1,j)=0,j>0,\;
\Pi(n,0)=0,n\ge2.\]
Also
\[\Pi(2,1)=\Pi(3,1)=\Pi(4,1)=1.\]
Some values of $P$ and $\Pi$ is given in Tables \ref{T1} and
\ref{T2}. 
(We use in the tables and figures the subscript $\xxl$ to emphasise
that we deal with the linear case.)

\begin{table}
\begin{center}
\begin{tabular}{|c|ccccc|}
\hline
\setlength{\unitlength}{0.25cm}
\begin{picture}(2,2)
\put(2.0,0.6){$k$}
\put(-0.4,-0.1){$n$}
\drawline(-1,2)(2.5,-0.5)
\end{picture}
 &0 & 1 & 2 &3&4\\
\hline
1 &$1$ &$ 0$ &$0$ &$ 0$ &$ 0$\\
2 &$1$ &$ 0$ &$0$ &$ 0$ &$ 0$\\
3 &$2/3$ &$ 1/3$ &$0$ &$ 0$ &$ 0$\\
4 &$1/3$ &$ 2/3$ &$0$ &$ 0$ &$ 0$\\
5 &$2/15$ &$ 11/15$ &$2/15$ &$ 0$ &$ 0$\\
6 &$2/45$ &$ 26/45$ &$17/45$ &$ 0$ &$ 0$\\
7 &$4/315$ &$ 38/105$ &$4/7$ &$ 17/315$ &$ 0$\\
\hline
\end{tabular}
\end{center}
\caption{$ P\xl(n,k)$}
\label{T1}
\end{table}

\begin{table}
\begin{center}
\begin{tabular}{|c|cccc|}
\hline
\setlength{\unitlength}{0.25cm}
\begin{picture}(2,2)
\put(2.0,0.6){$j$}
\put(-0.4,-0.1){$n$}
\drawline(-1,2)(2.5,-0.5)
\end{picture}
 &0 & 1 & 2 &3\\
\hline
0 &$1$ &$ 0$ &$0$ &$ 0$ \\
1 &$1$ &$ 0$ &$0$ &$ 0$ \\
2 &$0$ &$ 1$ &$0$ &$ 0$ \\
3 &$0$ &$ 1$ &$0$ &$ 0$ \\
4 &$0$ &$ 1$ &$0$ &$ 0$\\
5 &$0$ &$13/15$ &$2/15$ &$ 0$ \\
6 &$0$ &$ 28/45$ &$17/45$ &$ 0$ \\
7 &$0$ &$ 118/315$ &$197/315$ &$ 0$ \\
 $\cdot$ &    &         &            & \\
20 &0 &$ <10^{-7}$ &$\bullet$ &$ \bullet$ \\
\hline
\end{tabular}
\end{center}
\caption{$ \Pi\xl(n,j)$}
\label{T2}
\end{table}

Denoting the $j$th column of $\Pi$ by $\pi^{(j)}$, we have
\[\pi^{(j)}=P^{j-1} \pi^{(1)}.\]
For $j\ge2$, it suffices to consider $k\ge2$ in \eqref{E1}, so we need
only the matrix $(P(n,k))_{n,k\ge2}$. Since $P(n,k)=0$ if $k>(n-1)/2$,
this matrix is triangular, and so is $P^{j-1}$.

But $\pi^{(1)}(n)<10^{-7}$, $n>20$, so numerically, the significant
columns of $P^{j-1}$ are the first $20$ columns. Also, we see the
importance of the initial first column of $\Pi$. Moreover, for $n>75$,
$P(n,k)$ is indistinguishable from the Gaussian limit. So we have used
the expansion of the GF (\ref{E0}) for $n\leq 75$ and the Gaussian
limit afterwards in our numerical calculations. 
Of course we have
\[x(n)=\sum_{j=0}^\II \Pi(n,j)j,\] 
and
\[x(n)=1+\sum_{k=2}^\II P(n,k)x(k), \qquad n\ge2.\]

\begin{rem}
Another approach could be the following:
Let
\bals
I(k)&:=\kl \mbox{ one of the phases has $k$ players }\kr, \\
I(j,k)&:=\kl  \mbox{ phase $j$  has $k$ players }\kr,\\
Q(k)&:=\Pr[I(k)=1],\\
R(j,k)&:=\Pr[I(j,k)=1],\\
Q(k)&=\Pr\Bigl[\bigvee _j I(j,k)=1\Bigr]=\sum_{j=1}^{n-k} R(j,k), \mbox{ as }
\Pr[I(j,k)\wedge I(i,k)]=0, \mbox{ if }i\neq j,\\ 
R(j,k)&=\sum_l R(j-1,l)P(l,k),\; j\ge1, \text{ and } R(0,k)=\delta_{kn}\\
x(n)&=\E\Bigl[\sum_1^n I(k)\Bigr]=\sum_1^n Q(k) .
\end{align*}
\end{rem}

\begin{figure}[ht]
	\center
		\includegraphics[width=0.45\textwidth,angle=270]{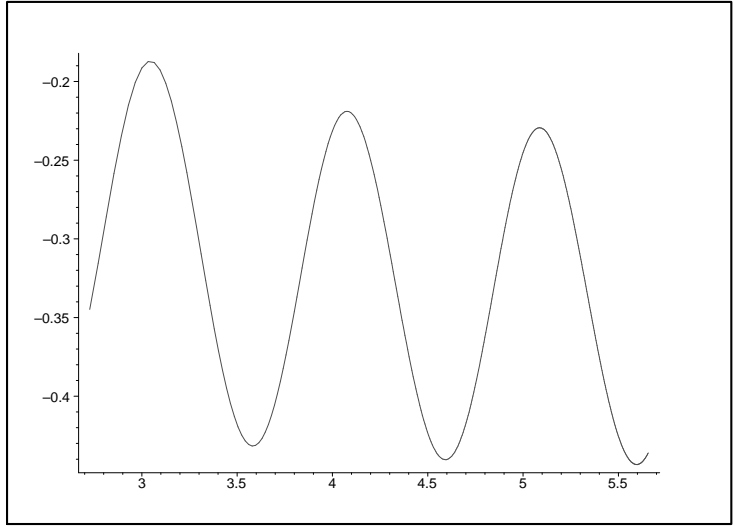}
		\medskip
		
	\caption{$x\xl(n)-\log_3 n $ versus $\log_3 n$, $n=50,\dots,500$  }
	\label{F1}
\end{figure}

A plot of $x(n)-\log_3 n $ versus $\log_3 n$ is given in Figure
\ref{F1}
for $n=50,\dots,500$. 
The oscillations expected from \eqref{EXn} are clear.

\begin{figure}[ht]
	\center
		\includegraphics[width=0.45\textwidth,angle=270]{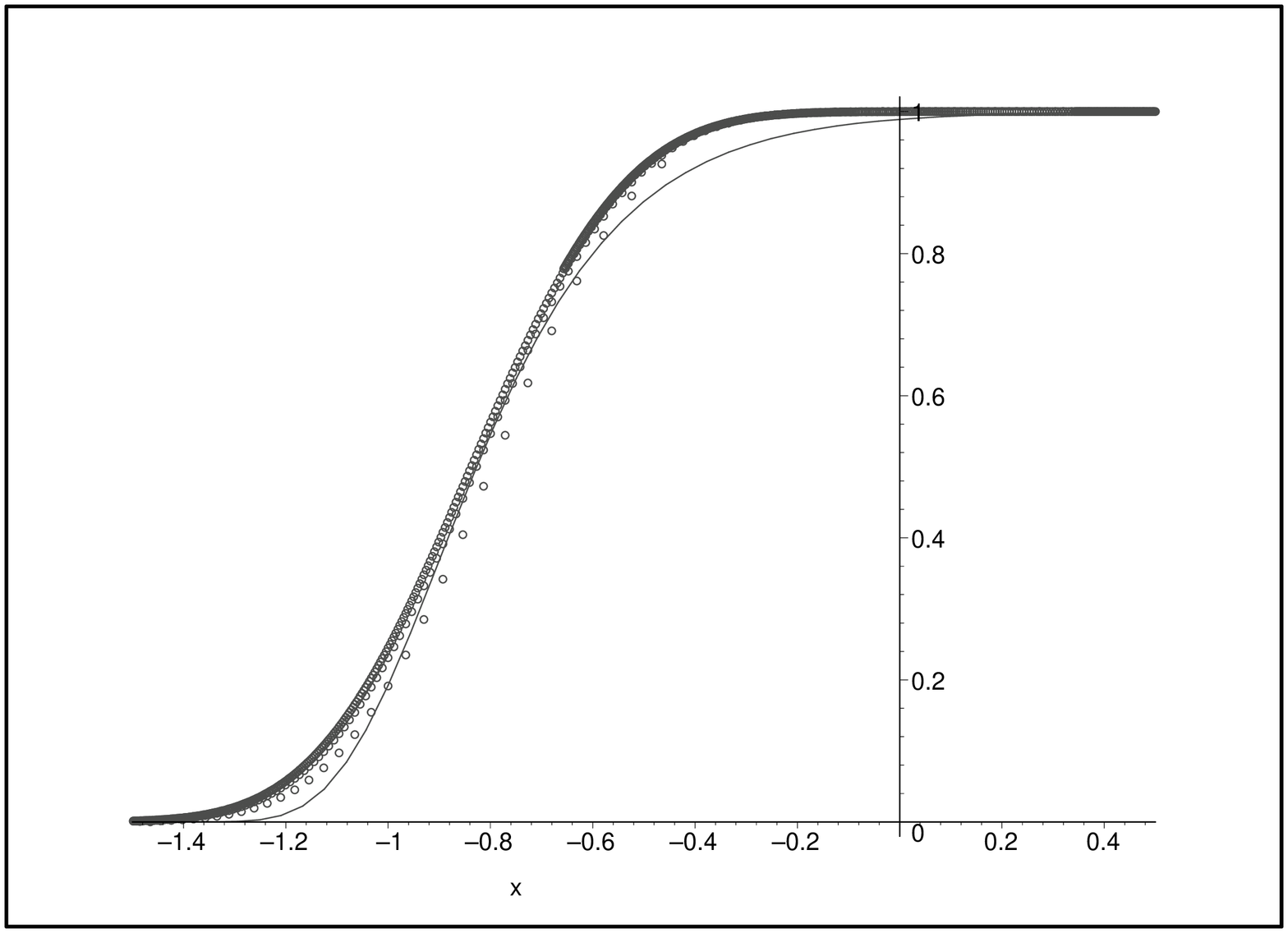}
		\medskip
   \begin{flushleft}
       \hspace{3.1cm} $\circ$\, : observed \\
       \hspace{3cm} --- : Gumbel distribution
   \end{flushleft}		
	\caption{$\La\xl(n,j)=\Pr(X_n\le j)$ versus $j-\log_3 n$,
approximating $F\xl(x)$, $n=20,\dots,500$}
	\label{F2}
\end{figure}

Recall that according to \refT{Th1}, there  exists a limiting
distribution function $F(x)=F\xl(x)$ (in a certain sense) for $X_n$. 
In Figure \ref{F2}, we approximate this distribution function $F(x)$ 
by plotting $\La(n,j)=\Pr(X_n\le j)$ against $j-\log_3n$ for
$n=20,\dots,500$, cf.\ \eqref{t1a}.
We have also plotted a scaled 
Gumbel distribution; the fit is bad.  

\begin{figure}[ht]
	\center
		\includegraphics[width=0.45\textwidth,angle=270]{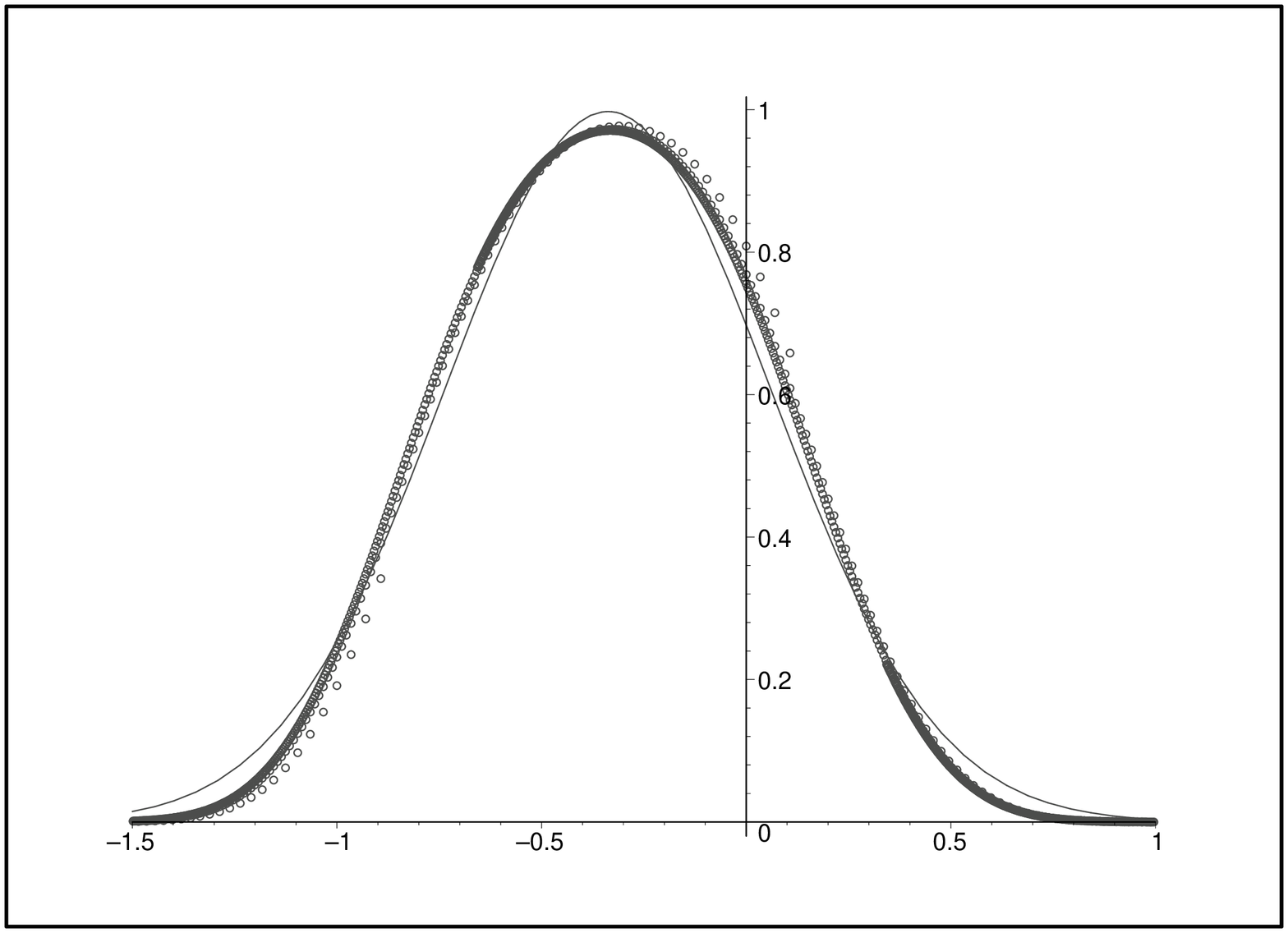}
		\medskip
   \begin{flushleft}
     \hspace{3.1cm} $\circ$\, : observed \\
     \hspace{3cm} --- : Gaussian distribution
   \end{flushleft}				
	\caption{$\Pi\xl(n,j)=\Pr(X_n=j)$ versus $j-\log_3 n$,
approximating $\DF\xl(x)$, $n=150,\dots,500$}
	\label{F3}
\end{figure}

Similarly, in Figure \ref{F3} we show the probability
$\Pi(n,j)$, $n=150,\dots,500$, plotted against $j-\log_3 n$. The fit with a
Gaussian distribution is equally bad.

\begin{figure}[ht]
	\center
		\includegraphics[width=0.45\textwidth,angle=270]{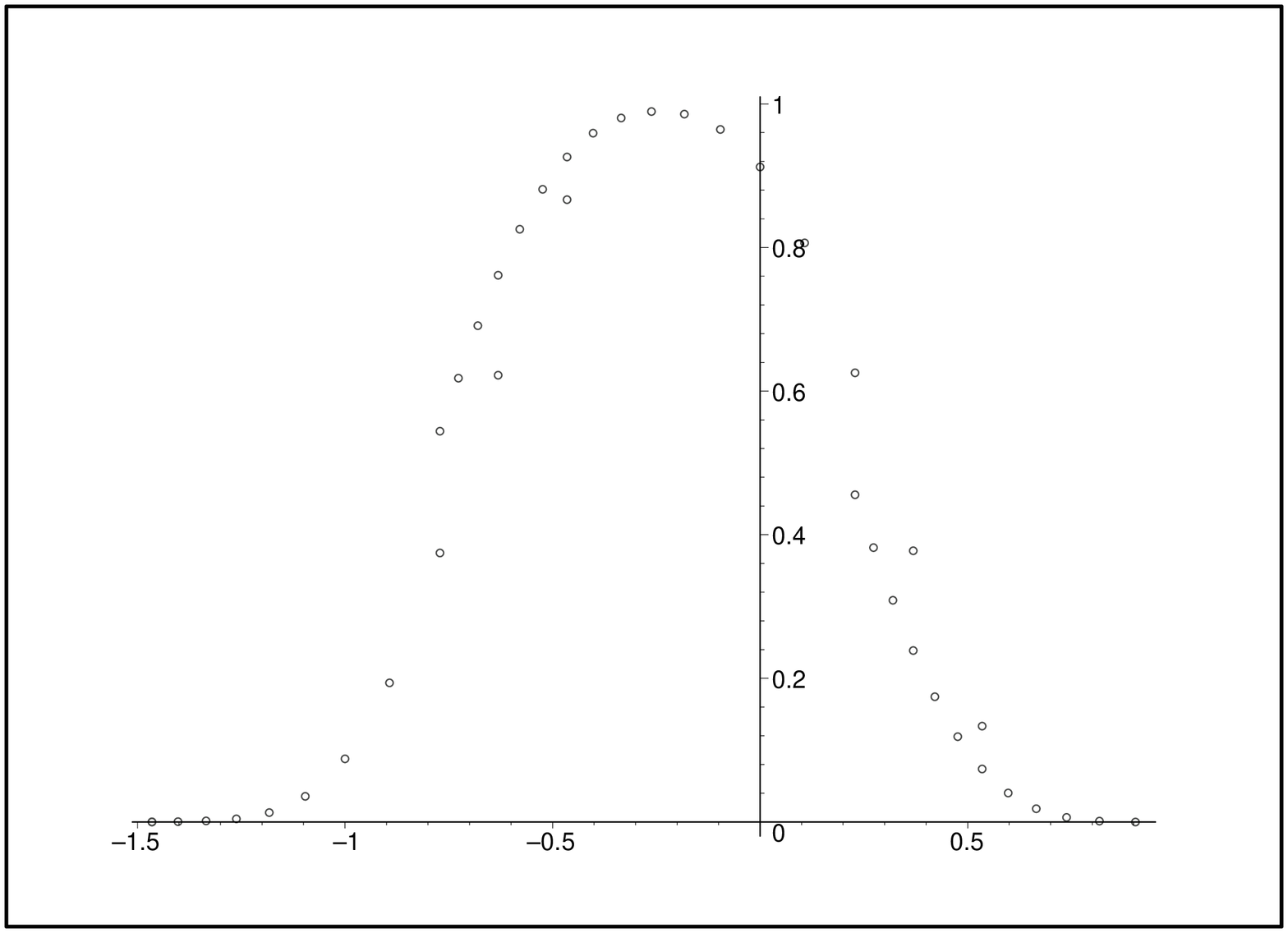}
		\medskip
		
	\caption{$\Pi\xl(n,j)$ versus $j-\log_3 n$, $n=1,\dots,40$}
	\label{F4}
\end{figure}

The few scattered points of both figures are actually due to small $n$
and the
propagation of the more erratic behaviour for $n=1,\dots,40$ shown in
Figure \ref{F4}.

\begin{figure}[ht]
	\center
		\includegraphics[width=0.45\textwidth,angle=270]{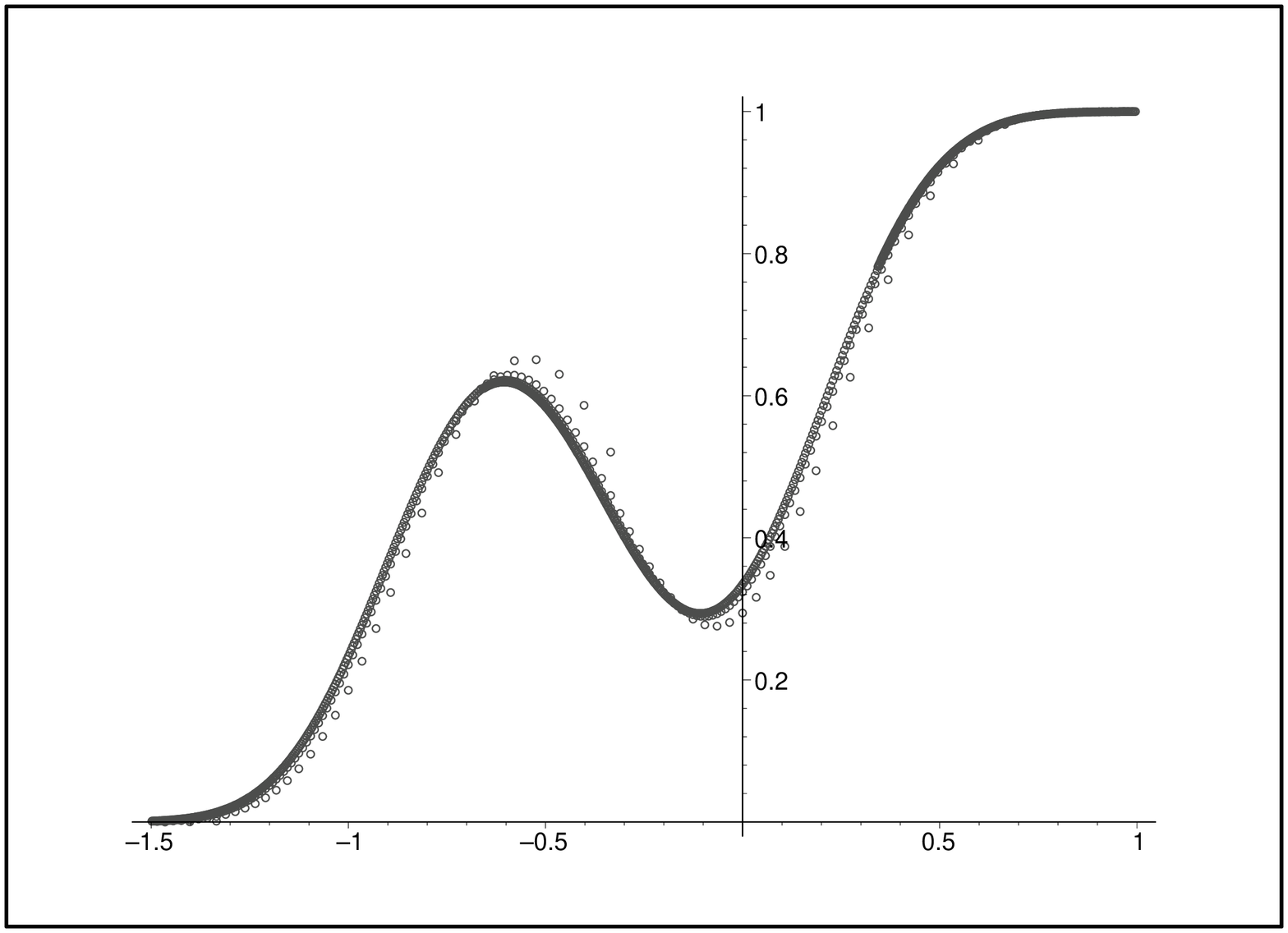}
		\medskip
		
\caption{$\La(n,j)$ versus $j-\log_3 n$, other initialization,
  $n=5,\dots,500$  } 
	\label{F5}
\end{figure}

\begin{figure}[ht]
	\center
		\includegraphics[width=0.45\textwidth,angle=270]{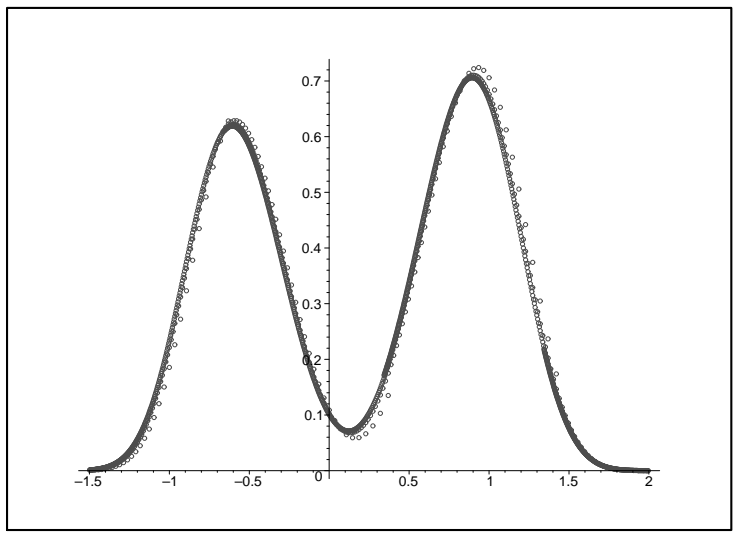}
		\medskip
		
\caption{$\Pi(n,j)$ versus $j-\log_3 n$, other initialization,
  $n=5,\dots,500$  } 
	\label{F6}
\end{figure}

\begin{figure}[ht]
	\center
		\includegraphics[width=0.45\textwidth,angle=270]{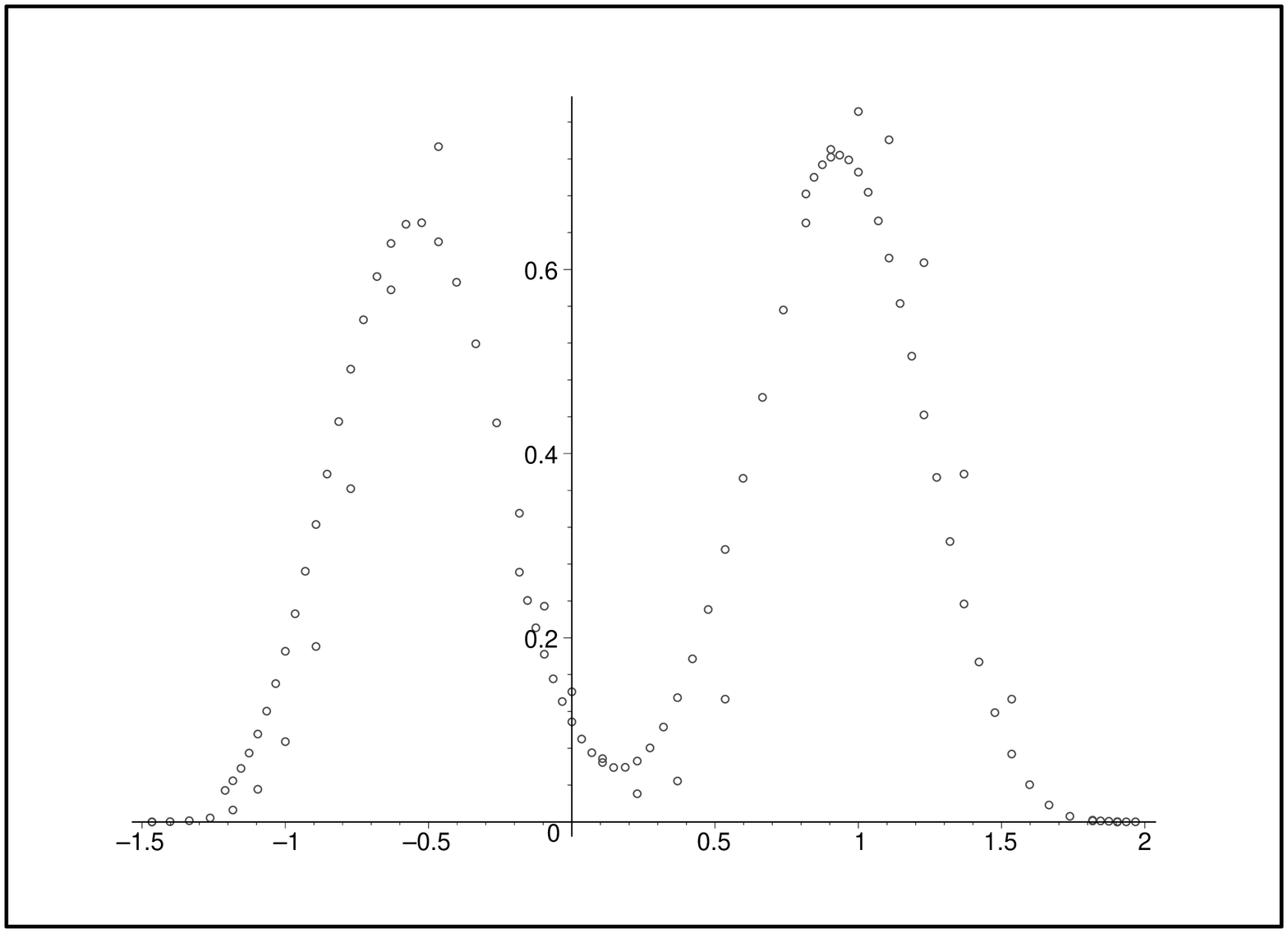}
		\medskip
		
	\caption{$\Pi(n,j)$ versus $j-\log_3 n$, $n=1,\dots,100$, other
	initialization  } 
	\label{F7}
\end{figure}

So we observe the following facts:
\begin {romenumerate}
\item  A first regime ($n=1,\dots,40$) create some scattered points
  which almost look like two distributions.
\item Between $n=40$ and $n\sim 75$ , a limiting distribution is
  attained; $\Pi(n,j)$ is concentrated on $j=\log_3 n +\BO(1)$, 
\item For $n>75$, the limiting Gaussian for $P$ with its narrow
 ($\sqrt{n}$) dispersion, intuitively induces, with (\ref{E1}), a
 propagation, with some smoothing, of the previous distribution. We
 attain the  limiting distribution $F(x)$ given by Theorem
 \ref{Th1}. 

\item 
At most two  values carry the main part of the
probability mass $\Pi(n,j)$. 
This is clear from the  observed range of $F(x)$ in Figures \ref{F2}
and \ref{F3}.
\item $P$ is triangular, and so is $P^j$. Also, $P^j(n,k)=\Theta(1)$, with
  $k=\BO(1)$, only if $j=\log_3 n +\BO(1)$.
\item  As $F(x)$ is absolutely continuous, we can derive, as in \cite{LP04},
  modulo some uniform integrability conditions, all (periodic) moments
  of $X_n$, in particular $x(n)$. 
\item \label{r41initial}
The effect of \emph{initial values} is now clear. 
To illustrate
  this we have changed  to 
$\Pi(0,1)=\Pi(1,0)=1$, which means that we add a cost $1$ for the
  extra selection required when the algorithm terminates with no
  element left. This leads to  Table \ref{T4}. The 
  equivalent of Figures \ref{F2}, \ref{F3} and \ref{F4}  is given in
  Figures \ref{F5}, \ref{F6} and \ref{F7}. 
Note that $X_n$ no longer is stochastically monotone in $n$; we have
by definition $X_0=1>0=X_1$, and Table \ref{T4} shows other examples
of non-monotonicity for small $n$. Moreover, Figure \ref{F5} shows
that the non-monotonicity persists for large $n$; we clearly have
convergence to a limit function, $G(x)$ say, but the limit is not
monotone and thus not a distribution function as in Figure \ref{F3}
and, more generally, in Theorem \ref{Th1}. 
\end {romenumerate}

\begin{rem}\label{Rnonmonotone} 
Note that the example in \ref{r41initial} and Figure \ref{F5}
does not contradict \refT{Th1} because $X_n$ now is defined with other
initial values than in \refT{Th1}. Nevertheless, it is a warning that
monotonicity of the limit should not be taken for granted in cases
such as \refE{Ebias} with $p<1/2$ where the monotonicity assumption of
\refT{Th1} is not satisfied.
\end{rem}

\begin{table}
\begin{center}
\begin{tabular}{|c|cccc|}
\hline
\setlength{\unitlength}{0.25cm}
\begin{picture}(2,2)
\put(2.0,0.6){$j$}
\put(-0.4,-0.1){$n$}
\drawline(-1,2)(2.5,-0.5)
\end{picture}
 &0 & 1 & 2 &3\\
\hline
0 & $0$ &$1$ &$0$ &$ 0$ \\
1 &$1$ &$0$ &$0$ &$ 0$ \\
2 &$0$ &$0$ &$1$ &$ 0$ \\
3 &$0$ &$1/3$ &$2/3$ &$ 0$ \\
4 &$0$ &$2/3$ &$1/3$ &$ 0$\\
5 &$0$ &$ 11/15$ &$2/15$ &$2/15$ \\
 $\cdot$ &    &         &            & \\
\hline
\end{tabular}
\end{center}
\caption{$ \Pi(n,j)$, other initialization}
\label{T4}
\end{table}

\subsection{Periodicities}\label{SPer}

Let $\psi(\a)\=\int e^{\a x}f(x)\,dx$ be the Laplace transform of the
limiting distribution $F$.
(We do not know whether $\psi(\a)$ exists
in general, although we conjecture so, but we really only need it for
imaginary $\a$, \ie, the characteristic function of $F$.)
Similarly, let 
$\tpsi(\a)\=\int e^{\a x}\DF(x)\,dx$ be the Laplace transform of $\DF$.
Since $\DF(x)\=F(x)-F(x-1)=\int_0^1f(x-t)\,dt$, $\DF=f*\chi_{[0,1]}$,
and thus, since $\chi_{[0,1]}$ has Laplace transform $(e^\a-1)/\a$,
\begin{equation*}
  \tpsi(\a)
=
\frac{e^\a-1}{\a}\psi(\a).
\end{equation*}
With the usual machinery (see \cite{LP04} or \cite[Theorem 2.3]{JA04}), 
we  obtain from \refT{Th1} and \refR{RZ}, assuming some technical
conditions that are very likely to hold but not rigorously verified,
\begin{align}
  \label{lap1}
x(n):=\E X_n&= \log_3n+\tilde{m}_1+w_1(\log_3n)+o(1), 
\intertext{where}
\tilde{m}_1&:=\tpsi'(0)
=\psi'(0)+\tfrac12, \label{lap2}\\
w_1(x)&:=\sum_{l\neq 0}\tpsi'(2\pi l\ii)e^{2\pi l\ii x}
=\sum_{l\neq0} \frac{\psi(2\pi l\ii)}{2\pi l \ii}e^{2\pi l\ii x}
.\label{lap3}
\end{align}

\begin{figure}[ht]
	\center
		\includegraphics[width=0.45\textwidth,angle=270]{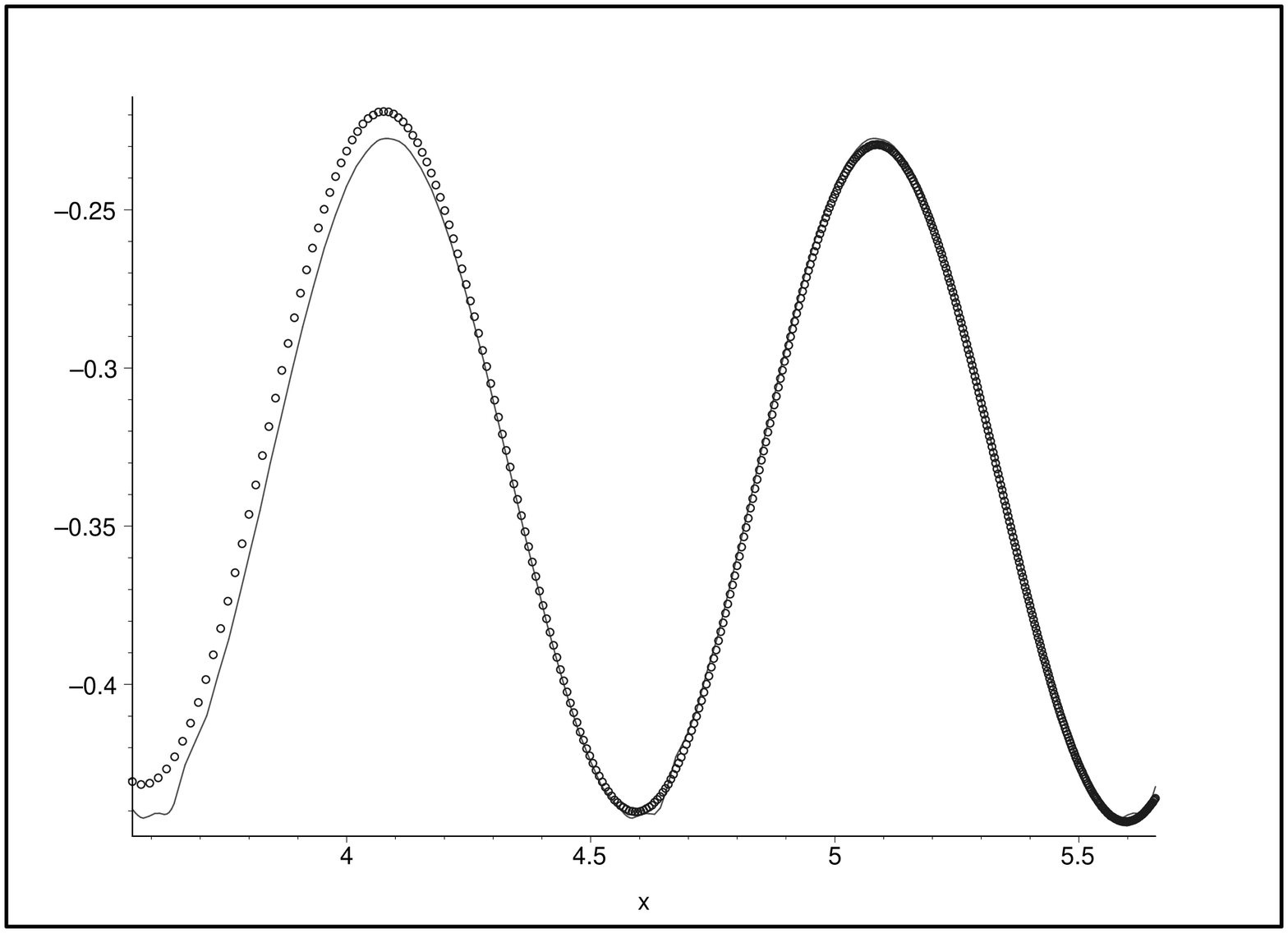}
		\medskip
		
\caption{observed  $x(n)-\log_3 n$ ($\circ$) and  computed with
(\ref{lap1}) (line) periodicities versus $\log_3 n$ (linear case),
  $n=50,\dots,500$ }  
	\label{F8}
\end{figure}

With the observed values of $\Pi(n,j)$ (see Figure \ref{F3}), we have
computed the Laplace transform numerically (with a variable step
Euler--MacLaurin) as follows. Assume that we have $N$ computed values
$\{\Pi(n_i,j_i):i=1, \ldots, N\}$. Setting $x=j-\log n$, this gives
$\{\Pi(n_i,x_i):i=1, \ldots, N\}$. Sorting wrt $x_i$, we write this as
$\{\Pi(n_k,y_k):y_k\leq y_{k+1},\,k=1, \ldots, N\}$. Construct a numerical
Laplace transform 
\[\tpsi(\al)=\sum e^{\al y_k}\Pi(n_k,y_k)(y_{k+1}-y_{k-1})/2 .\]
Using this numerically computed $\psi$ in \eqref{lap1}--\eqref{lap3}, 
we compute $\tilde{m}_1+w_1(x)$, 
which fits quite well with the observed periodicities 
of $x(n)-\log_3 n$
in
Figure \ref{F1}; the comparison is given 
in Figure \ref{F8}. 

\section{Second variation of the Franklin leader election
  algorithm. The circular  case}\label{SS2} 

If we denote by $P\xc(n,k)$ the distribution of the number of peaks in
the circular case and by  $P\xl(n,k)$ the distribution in the linear
case, we know, by  \refE{EX39} that $P\xc(n,k)=P\xl(n-1,k-1)$. It is
easy to check that this leads to, for $n\ge3$ and $n\ge5$, respectively,
\[\M(n)=n/3,\quad\V(n)=2n/45,\]
which also is easy to see probabilistically, by writing $Y_n$ as the
sum of the $n$ indicators 
$\kl \mbox{player $i$ is a peak}\kr$, and noting that indicators with
distance at least 3 are independent.)

The initial values are now given by
\[P(1,1)=1,\;P(2,1)=1,\;P(3,1)=1,\;
\Pi(0,0)=1,\;\Pi(1,0)=1,\;\Pi(2,1)=1,\;\Pi(3,1)=1.\]
The corresponding pictures are given in Tables \ref{T5} and  \ref{T6}.
(We use a subscript $\xxc$ for the circular  case.)

\begin{table}
\begin{center}
\begin{tabular}{|c|ccccc|}
\hline
\setlength{\unitlength}{0.25cm}
\begin{picture}(2,2)
\put(2.0,0.6){$k$}
\put(-0.4,-0.1){$n$}
\drawline(-1,2)(2.5,-0.5)
\end{picture}
 &0 & 1 & 2 &3&4\\
\hline
1 &$0$ &$ 1$ &$0$ &$ 0$ &$ 0$\\
2 &$0$ &$ 1$ &$0$ &$ 0$ &$ 0$\\
3 &$0$ &$ 1$ &$0$ &$ 0$ &$ 0$\\
4 &$0$ &$ 2/3$ &$1/3$ &$ 0$ &$ 0$\\
5 &$0$ &$ 1/3$ &$2/3$ &$ 0$ &$ 0$\\
6 &$0$ &$2/15$ &$11/15$ &$ 2/15$ &$ 0$\\
7 &$0$ &$ 2/45$ &$26/45$ &$ 17/45$ &$ 0$\\
\hline
\end{tabular}
\end{center}
\caption{$ P\xc(n,k)$}
\label{T5}
\end{table}

\begin{table}
\begin{center}
\begin{tabular}{|c|cccc|}
\hline
\setlength{\unitlength}{0.25cm}
\begin{picture}(2,2)
\put(2.0,0.6){$j$}
\put(-0.4,-0.1){$n$}
\drawline(-1,2)(2.5,-0.5)
\end{picture}
 &0 & 1 & 2 &3\\
\hline
0 &$1$ &$ 0$ &$0$ &$ 0$ \\
1 &$1$ &$ 0$ &$0$ &$ 0$ \\
2 &$0$ &$ 1$ &$0$ &$ 0$ \\
3 &$0$ &$ 1$ &$0$ &$ 0$ \\
4 &$0$ &$ 2/3$ &$1/3$ &$ 0$\\
5 &$0$ &$ 1/3$ &$2/3$ &$ 0$\\
6 &$0$ &$ 2/15$ &$13/15$ &$ 0$\\
7 &$0$ &$ 2/45$ &$43/45$ &$ 0$\\
 $\cdot$ &    &         &            & \\
20 &0 &$ <10^{-7}$ &$\bullet$ &$ \bullet$ \\
\hline
\end{tabular}
\end{center}
\caption{$ \Pi\xc(n,j)$}
\label{T6}
\end{table}

\begin{figure}[ht]
	\center
		\includegraphics[width=0.45\textwidth,angle=270]{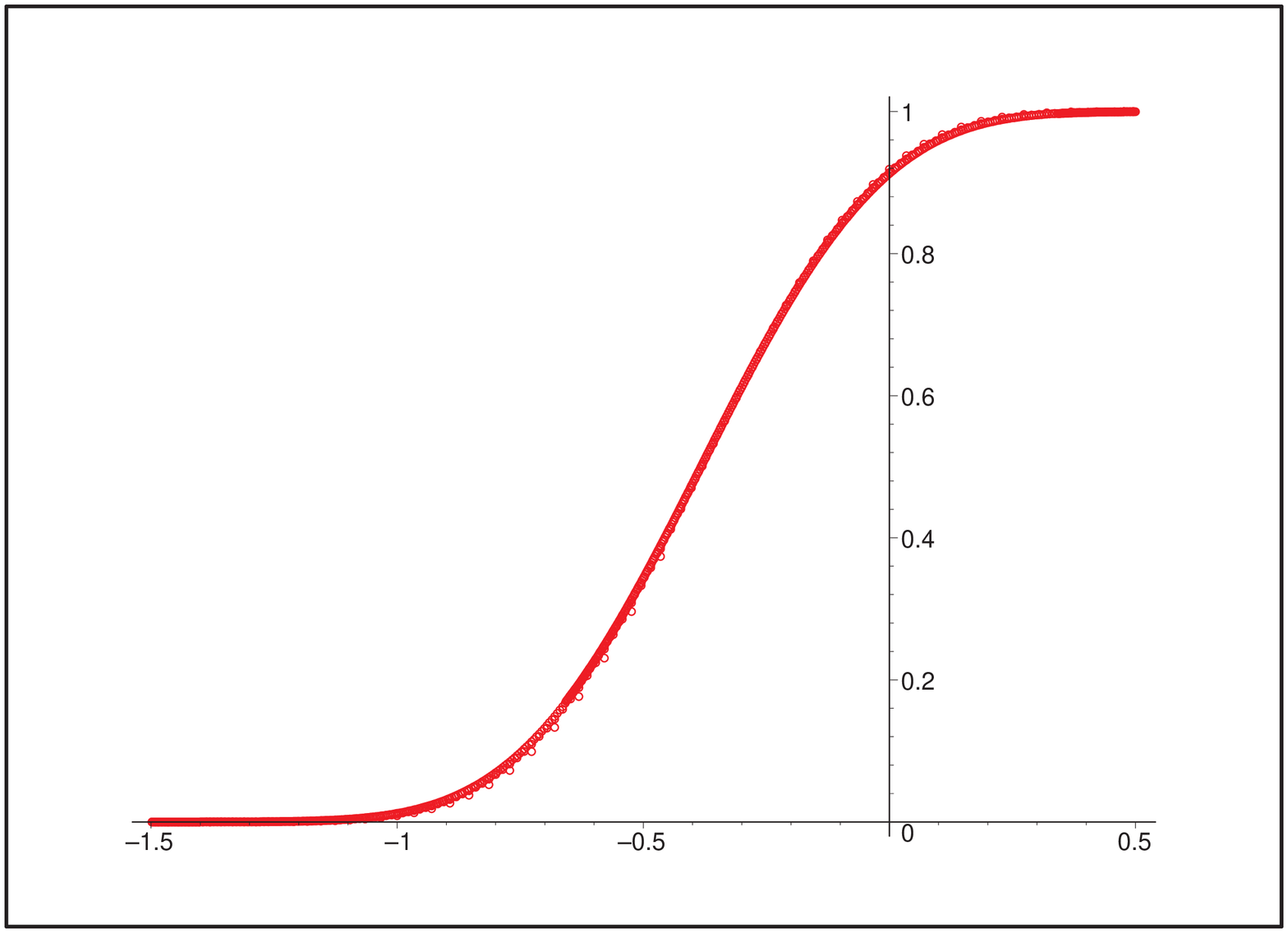}
		\medskip
   
\caption{$\La\xc(n,j)$ versus $j-\log_3 n$, approximating $F\xc(x)$,
  $n=5,\dots,500$} 
	\label{F9}
\end{figure}

\begin{figure}[ht]
	\center
		\includegraphics[width=0.45\textwidth,angle=270]{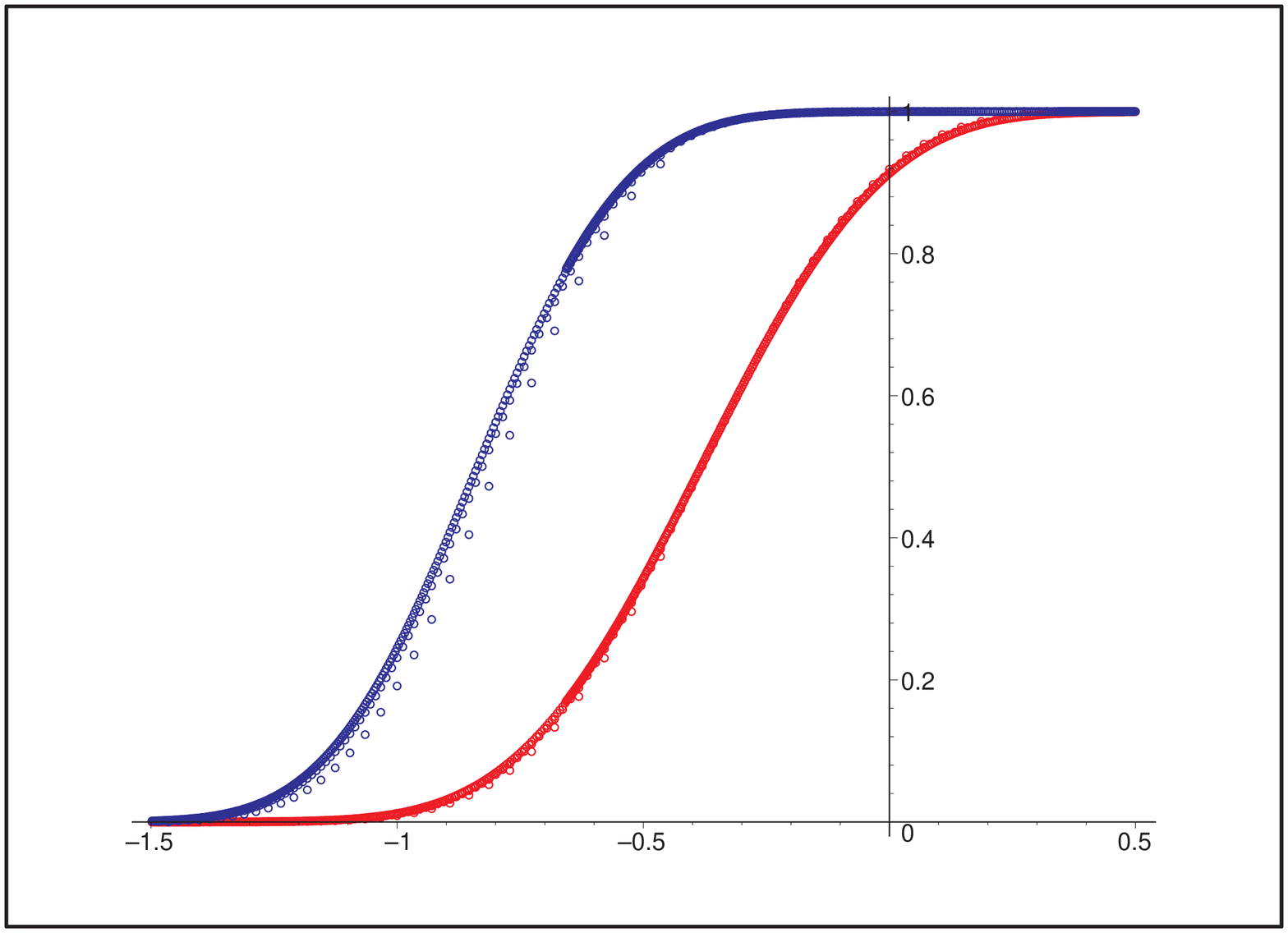}
		\medskip
   
\caption{A comparison of $\La\xc(n,j)$ (right curve)  with
  $\La\xl(n,j)$  (left curve)} 
	\label{F10}
\end{figure}

A plot of $\La\xc(n,j)$ versus $j-\log_3 n$  for $n=20,\dots,500$  is
given in Figure 
\ref{F9}. We see that there are fewer scattered points than in the
linear case. Let us mention that the fits with Gumbel or Gaussian are
equally bad. A comparison of $\La\xc(n,j)$ with $\La\xl(n,j)$ is given
in Figure \ref{F10}. No numerical relation exists between the two
distributions.

\begin{figure}[ht]
	\center
		\includegraphics[width=0.45\textwidth,angle=270]{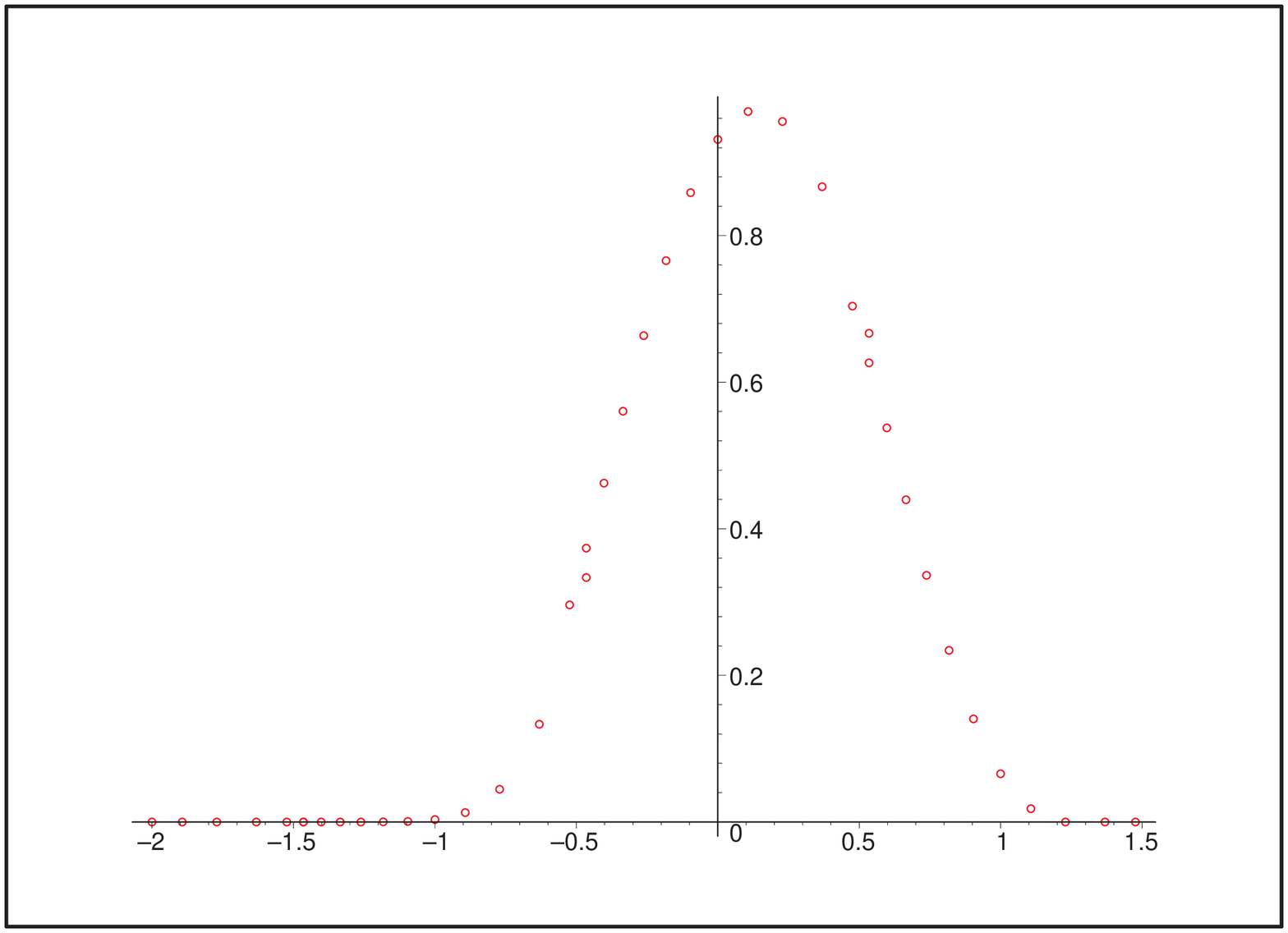}
		\medskip
	\caption{$\Pi\xc(n,j)$ versus $j-\log_3 n$, $n=1,\dots,40$  }
	\label{F11}
\end{figure}

$\Pi\xc(n,j)$ versus $j-\log_3 n$, $n=1,\dots,40$, is plotted in
Figure \ref{F11}. 
These initial points are now scattered less than in the linear case.

\begin{figure}[ht]
	\center
		\includegraphics[width=0.45\textwidth,angle=270]{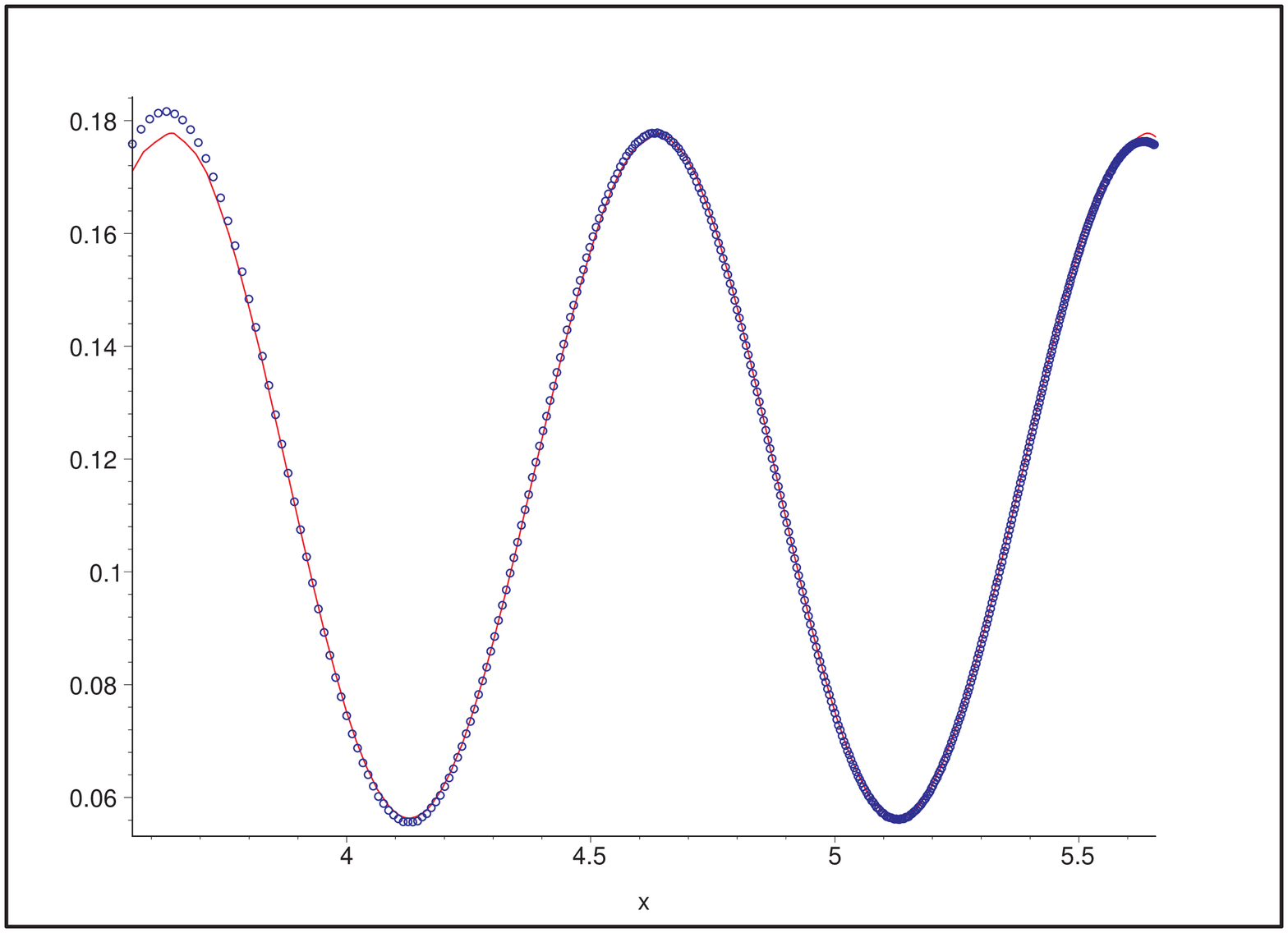}
		\medskip
		
\caption{observed  $x(n)-\log_3 n$ ($\circ$) and  computed
with (\ref{lap1}) (line) periodicities versus $\log_3 n$ 
(circular case), $n=50,\dots,500$ } 
		\label{F12}
\end{figure}

The observed versus computed  periodicities are given in Figure \ref{F12}.

In conclusion, apart from numerical differences, the behaviour of our
two variations are quite similar. 

Note that the mean number of needed messages is asymptotically $2n
\log_3(n)$, as we use $2n$ messages per round. Franklin
\cite{FR82}
gives an upper bound $2n \log_2(n)$.

\bibliographystyle{plain} 
\bibliography{biblio-sj211}

\end{document}